\documentclass[11pt]{article}
\usepackage[hmargin=2cm,vmargin=1.5cm]{geometry}
\usepackage{graphicx}
\usepackage{amsmath}
\usepackage{enumerate}

\begin{document}
\title{Time-of-arrival probabilities for general particle detectors}
\author {Charis Anastopoulos\footnote{anastop@physics.upatras.gr} \\
 {\small Department of Physics, University of Patras, 26500 Greece} \\
 and  \\ Ntina Savvidou\footnote{ntina@imperial.ac.uk} \\
  {\small  Theoretical Physics Group, Imperial College, SW7 2BZ,
London, UK} }

\maketitle
\begin{abstract}
We develop a general framework for the construction of probabilities for the time of arrival in quantum systems. The time of arrival is identified with the time instant when a transition in the detector's degrees of freedom takes place. Thus, its definition is embedded within the larger issue of defining probabilities with respect to time for general quantum transitions.   The key point in our analysis is that we manage to reduce the problem of defining a quantum time observable to a mathematical model where time is associated to a transition from a subspace of the Hilbert space of the total system to its complementary subspace. This property makes it possible to derive a general expression for the probability for the time of transition, valid for any quantum system, with the only requirement that the time of transition is correlated with a definite macroscopic record.

The framework developed here allows for the consideration of any experimental configuration for the measurement of the time of arrival, and it also applies to relativistic systems with interactions described by quantum field theory. We use the method in order to describe time-of-arrival measurements in high-energy particle reactions and for a rigorous derivation of the time-integrated probabilities in particle oscillations.
\end{abstract}
\section{Introduction}

The time-of-arrival issue \cite{ML, ToAbooks}, in its simplest form, amounts to the following problem. One considers an initial wave function $|\psi_0 \rangle$ for a particle centered around $x = 0$ and with a positive mean momentum. The question is to find the probability $P(t)dt$ that the particle is detected at distance $x = L$ at some moment between $t$ and $t+\delta t$. The issue is important in the foundations of quantum mechanics, in relation to the role of time in the theory, but also because of the  possibility of measuring time-of-arrival probability distributions.

In this paper, we develop a general framework for the construction of probabilities for the time-of-arrival in quantum systems. The key idea is the inclusion of the measuring apparatus in the quantum description; hence, the time of arrival is defined as a coarse-grained observable associated to the macroscopic records of the apparatus that correspond to a particle's detection. Our framework allows for the consideration of any experimental configuration for the measurement of the time of arrival. In particular, our method applies to relativistic systems with interactions described by quantum field theory, therefore it is particularly suitable for time-of-arrival measurements in high-energy physics.

A common conclusion emerging from the different approaches to the topic is that the definition of probabilities for the time of arrival strongly depends on the specific experimental set-up through which the time of arrival is determined. The latter observation is the starting point of our method. Rather than attempting to construct a time-of-arrival probability for the properties of the microscopic quantum particle, we consider the larger system including the microscopic particle and a measurement apparatus. In the combined system, the time of arrival is associated to a definite macroscopic record of particle detection, defined in terms of the apparatus' degrees of freedom. We can then treat the time of arrival as a quasi-classical variable, and construct the relevant probabilities using standard expressions from the decoherent histories approach to quantum theory \cite{Omn1, Omn2, Gri, GeHa, hartlelo}.

 We treat the time of arrival  as a special instance of the more general notion of the time associated to a transition in a quantum system---here the transition refers to the degrees of freedom of the measuring apparatus. Thus, in order to describe the time of arrival, we first derive an operator expression for the probability associated to transition times in general quantum systems. This probability distribution depends only on the Hamiltonian, the projection operators that define the transition and the initial state of the system. The derivation requires no assumptions about specific properties of the physical system; only that the time of transition is associated to a macroscopic record of observation. Hence, the most important contribution of this paper is that it provides a general methodology for the determination of the time-of-arrival probabilities associated to an experiment.

\subsection{The time-of-arrival issue}

In quantum theory, probability distributions for observables are   constructed using the Born rule, or its generalizations. However, the Born rule does not directly apply to the time-of-arrival problem because there is no self-adjoint operator for time in quantum mechanics \cite{PaEn, UnWa89}. The time  $t$  appearing in Schr\"odinger's equation is an external parameter and not an observable. This implies that the squared modulus of the time-evolved wave-function $|\psi(x, t)|$ is not a density with respect to $t$, and, hence, it cannot serve as a definition for the required probabilities \cite{AnSav06}.

The time-of-arrival problem is a special case of passing from the classical description of a magnitude to its associated quantum one. In this case,
the methodologies of the classical theory does not generalize to the quantum theory. In classical probability theory, time-of-arrival probability distributions are defined in terms of probability currents. For a non-relativistic particle of mass $m$, the probability current associated to Schr\"odinger's equation is
\begin{eqnarray}
J(L,t) = -\frac{i}{2m} \langle \psi_t|\hat{p} \delta(\hat{x}-L)+ \delta(\hat{x}-L) \hat{p}|\psi_t\rangle, \label{current0}
\end{eqnarray}
where $|\psi_t \rangle = e^{-i\hat{H}t}|\psi_0\rangle $ and $\hat{H}$ is the free particle's Hamiltonian. However, the current Eq. (\ref{current0}) gives rise to {\em negative} probabilities for initial states involving superpositions of different momenta \cite{current}.

The time-of-arrival issue is of significant interest because

\begin{itemize}

\item it provides a simple set-up for exploring different ideas about one of the major foundational issues in quantum mechanics, namely, the role of time in the theory. This issue is particularly pertinent on the fields of quantum
 gravity and quantum cosmology: the necessity of a reconciliation between the quantum notion of time as an external parameter to a physical system and the dynamical notion of time in general relativity generates the so-called problem of time in quantum gravity \cite{probtime}. In particular, ideas about the role of time can be applied both to the time of arrival problem and in quantum cosmology \cite{hartlelo, hal09, AnSav00}.

\item it is  a prototype for other physical problems in quantum theory that involve the definition of probabilities with respect to time.  Examples include the study of tunneling time \cite{tunnel1, AnSav08a}, that is, of the time that it takes a particle to cross a classically forbidden region, and the construction of well-defined probabilities for non-exponential decays  \cite{decay, AnSav08b}. Moreover, the rapid growth of quantum information theory brings into the forefront novel issues, such how entanglement is manifested at the level of probabilities for time. Conversely, the analogy with photo-detection theory, where temporal correlations provide significant information about the electromagnetic-field state \cite{Gla, Dav, AnSav11}, strongly suggests that  probabilities and correlations with respect to time can provide novel generalized criteria for entanglement.

\item time-of-arrival probability distributions can in principle be experimentally measured \cite{exp} allowing for a comparison between different theoretical predictions. The recent OPERA and ICARUS experiments have determined time-of-arrival probability distribution in the context of neutrino physics \cite{opera, icarus}. Furthermore, the theory describing particle oscillations \cite{GePa, Ponte} (neutrinos and neutral bosons) relies implicitly on the notion of the time of arrival \cite{zra}; the quantity relevant to the experiments is the total probability of particle detection integrated over all times of arrival $t$ of a particle at the detector.

\end{itemize}
  Currently,  there exist several different approaches to the time-of-arrival problem.

The main limitation of existing approaches is the lack of generality. There is no precise, algorithmic procedure allowing for the derivation of the time-of-arrival probabilities for any specific method of particle detection. Moreover, investigations are mainly restricted to non-relativistic quantum mechanics. They do not incorporate the quantum-field theoretic description of interactions, which would be necessary for the   study of time-of-arrival in high-energy physics.

For initial states sharply concentrated in momentum, all approaches  lead to probability distributions that are peaked around the classical value of the time of arrival. These probability distributions differ in their details, the differences being particularly pronounced for initial states with significant momentum spread.  In fact, time of arrival probabilities are strongly contextual, in the sense that they strongly depend on the experimental procedure through which the time of arrival is determined.

An axiomatic approach, developed by Kijowski, determines the probability distribution $P(L,t)$ for the time of arrival of non-relativistic particles by requiring Galilean-covariance and correspondence with the classical theory \cite{Kij}. The resulting expression is
\begin{eqnarray}
P(L,t) = \left| \int \frac{dp}{2 \pi} \sqrt{\frac{p}{m}} \tilde{\psi}_0(p) e^{ipL - i\frac{p^2}{2m}t}\right|^2, \label{Kij}
\end{eqnarray}
where $\tilde{\psi}_0$ is the initial state in the momentum representation. For wave functions with support on positive values of the momentum $p$, the probability density Eq. (\ref{Kij}) is normalized to unity for $t \in (-\infty, \infty)$. Other properties of the probability density Eq. (\ref{Kij}) are discussed in Ref. \cite{Kij2}, and generalizations in Ref. \cite{gener}.

Other approaches to the time of arrival include
\begin{itemize}
\item the use of complex potentials   modeling the absorption of the particle by a detector at $x = L$ \cite{Allcock, mpl99, Hall1, timen11};
\item the consideration of specific detector models \cite{exp, many, schul, hall2} or idealized clocks \cite{ha11};
 \item formulations within the decoherent histories approach to quantum mechanics \cite{hartlelo, yata, hartle2, miha, HaZa, HaYe};
 \item the analysis the time of first-crossing of $x = L$ for quantum mechanical paths. Such paths are    defined either using Feynman's prescription \cite{pathint}, or through Bohmian mechanics  \cite{bohmian}  or through phase space quasi-distributions \cite{wigner}.
\end{itemize}

\subsection{The proposed time-of-arrival algorithm}

 In this paper, we address the issue of constructing time-of-arrival probability distributions associated to {\em any} method of particle detection. To this end, we develop a novel method that is based on the following key ideas: (i) the inclusion of the detector degrees of freedom into the quantum description; (ii) the definition of the time-of-arrival as a coarse-grained quasi-classical variables associated to macroscopic records of particle detection, and (iii) the understanding of the time of arrival as a special case of the more general notion of transition-time, that applies to practically all quantum systems.

Our  method is  algorithmic: for each experimental set-up, one identifies the operators corresponding to the macroscopic records of particle detection, the total Hamiltonian (which includes the Hamiltonian for the microscopic particle, the Hamiltonian for the self-dynamics of the detector and an interaction term), and the initial state of the combined system. Once the above variables are determined, a unique expression for the time-of-arrival probability distribution follows. Moreover, the method involves no restrictions on properties of the operators through which the time-of-arrival probabilities are constructed; it applies to {\em any} quantum system, including relativistic systems interacting through quantum field theory. Hence, it is particularly suitable for the study of the time of arrival problem in  high-energy physics.

 We do not define the time of arrival as an intrinsic variable characterizing microscopic particles, but as a variable associated to degrees of freedom of macroscopic detectors. In particular, we identify the time of arrival with the instant $t$ that the apparatus "clicks", that is, with the reading of an external clock simultaneous to the creation of a macroscopic record  of  particle detection. A particle detection constitutes a macroscopic, irreversible amplification of a microscopic event; thus, the time-of-arrival $t$ is defined as a coarse-grained, quasi-classical variable associated with internal transitions of the detector\footnote{The time-of-arrival variable is coarse-grained in the sense that its value can only be ascertained with macroscopic accuracy, and it is quasi-classical in the sense that it is associated to definite macroscopic records. }. Probabilities for such quasi-classical variables are given by the
 the decoherent histories approach to quantum mechanics. We emphasize that in our approach, {\em a particle's time of arrival is defined only in presence of a definite fact of particle detection}.

The time-of-arrival issue then becomes a special case of the broader issue of  defining probabilities with respect to time for general quantum transitions; other special cases include the definition of decay probabilities in unstable systems \cite{decay},  and  the construction of coherence functions of the electromagnetic field in photo-detection \cite{Gla, Dav, AnSav11}. The key point in our analysis is that we manage to reduce the problem of defining a quantum time observable to a simple mathematical model where time is associated to a transition from   a subspace of the Hilbert space of the total system to its complementary subspace.
This property makes it possible to derive a general expression for the probability for the time of transition, valid for any quantum system, with the only requirement that the time of transition is correlated with a definite macroscopic record.

 Our method  involves three steps: (i) the derivation of an expression for the amplitudes associated to definite values for the  time $t$ that a specific transition takes place in a quantum system; (ii) the construction of the associated probabilities through the requirement that the time of transition $t$ is a coarse-grained quasi-classical variable; (iii) the specialization  to the time-of-arrival problem, by considering a system consisting of a microscopic particle interacting with a macroscopic apparatus at distance $L$ from the particle source. The result is a general expression for the  time-of-arrival probabilities. This expression involves certain operators that describe the properties of the particle detector, its interaction with the microscopic particle and the type of the recorded observables. These operators are determined by the physics of the time-of-arrival experiment in consideration; once they are specified, a unique expression for the time-of-arrival probabilities association to the experiment follows.

   Our main results are the following.
\begin{itemize}
\item We construct the time-of-arrival probabilities for three different models of particle detectors. We find that, in general, the time of arrival probabilities are strongly dependent on the physics of the detector, but there is an important regime where all information about the detector is encoded in a single function of the microscopic particle's momentum, the {\em absorption coefficient} $\alpha(p)$.
\item We identify a generalization of Kijowski's probability distribution that is valid for any dispersion relation for the microscopic particle, including relativistic ones. This probability distribution corresponds to the case of constant absorption coefficient.
\item We construct a general expression for the time-of-arrival probability in high-energy processes, in which the microscopic particle is detected through a reaction described by relativistic quantum field theory. In this context, our method leads to a general quantum-measurement-theoretic description of particle detectors in high-energy physics.
\item A non-trivial application of our formalism is the rigorous derivation of the time-integrated probabilities associated to particle-oscillation experiments. We obtain the standard oscillation formulas in a regime that corresponds to very short values of the decoherence times associated to the particle-detection process. Interestingly, we also find that in the regime of larger decoherence times a novel non-standard oscillation formula appears.
\end{itemize}

The present work originates from ideas in a broader program about the role of time in quantum theory \cite{Sav}. It
employs many concepts from the description of quantum measurements in the decoherent histories approach  \cite{Omn1, Gri, GeHa} and it has some similarities to the Davies-Srinivas photo-detection theory \cite{Dav}. Preliminary versions of the method \cite{AnSav06} have been employed for the description of tunneling time \cite{AnSav08a}, non-exponential decays \cite{AnSav08b}, and for the study of temporal correlations in particle detectors in relation to the Unruh effect \cite{AnSav11}.

The structure of this paper is the following. In Sec. 2, we derive a general formula for the probability distribution associated to the time of transition in any quantum system, and then we employ this formula in order to define probabilities associated to general time-of-arrival measurements. In Sec. 3, we study the time-of-arrival using idealized models for the particle detection process, we obtain a generalization of Kijowski's probability distribution, and we apply our formalism to time-of-arrival experiments in high energy physics. Sec. 4 contains a non-trivial application of our formalism, in the rigorous derivation of the time-integrated probability in particle-oscillation experiments.

\section{Derivation of a general formula for time-of-arrival probabilities}

In this section, we derive an expression for the time-of-arrival probabilities, applicable to any scheme of detection for the microscopic particle. To this end, we first construct probabilities associated to the time of transition for a general quantum  system; the time-of-arrival probabilities arise as a special case. In particular, in Sec. 2.1 we present the physical assumptions and the notations of the formalism; in Sec. 2.2 we construct the quantum amplitudes associated to a definite value of the time of transition; in Sec. 2.3 we construct the probabilities with respect to the time of transition $t$ by assuming that $t$ is a coarse-grained quasi-classical variable; and in Sec. 2.4 we consider the special case of time-of-arrival measurements where the transitions under consideration correspond to the recording of a particle event by a macroscopic detector.

\subsection{Preliminaries}

The key points of our derivation of the probabilities associated to the time of transition of a generic quantum system are the following.
\begin{enumerate}
\item Physical transitions in a quantum system are described  as  transitions between two complementary subspaces in the system's Hilbert space.

\item The time of a transition is defined in terms of a macroscopic record in an apparatus that is correlated to the microscopic transition event. Hence, the time of transition is a coarse-grained, decoherent observable, for which probabilities can be meaningfully defined.
\end{enumerate}

Point 1 is standard in ordinary quantum theory. For example,
\begin{itemize}{}
 \item the emission of a photon from an atom corresponds to a
transition from the one-dimensional subspace, defined by the
electromagnetic field vacuum, to the subspace of single-photon
states \cite{Dav};

\item a von Neumann measurement corresponds to  a transition from the subspace in which the pointer variable $\hat{X}$ takes its pre-measurement values, to a subspace  corresponding to possible measurement outcomes;

\item any particle reaction  can be described as a transition from the subspace of states associated to the initial particles to the subspace of the product particles.
\end{itemize}

Point 2  restricts the context of this study to observed transitions, i.e., to transitions associated to  definite macroscopic facts of observation. The existence of a macroscopic record necessitates that we include a measurement apparatus in the quantum description of the process.

We proceed to definitions of the quantities relevant to our construction of time-of-transition probabilities. We denote by  ${\cal H}$ be the
Hilbert space of the combined system, describing the degrees of freedom of both  the quantum  system under consideration and the macroscopic apparatus.

 We assume that
 ${\cal H}$ splits into two subspaces: ${\cal H} = {\cal
H}_+ \oplus {\cal H}_-$. The subspace ${\cal H}_+$ describes the accessible
states of the system given that a specific event is realized; the subspace ${\cal H}_-$ is the complement of ${\cal H}_+$. For example, if the quantum event under consideration is a detection of a particle by a macroscopic apparatus, the subspace ${\cal H}_+$ corresponds to all accessible states of the apparatus after a detection event has occurred.
We denote  the
projection operator onto ${\cal H}_+$ as $\hat{P}$ and
the projector onto ${\cal H}_-$ as $\hat{Q} := 1  - \hat{P}$.

Once the transition has taken place, it is possible to
measure the values of variables for the microscopic system through their correlation to a pointer variable of the measurement apparatus. We denote by $\hat{P}_\lambda$   projection operators (or, more generally, positive operators) corresponding to different values $\lambda$ of some physical magnitude that can be measured only if the quantum event under consideration has occurred. For example, when considering transitions associated to the detection of a particle, the projectors $\hat{P}_\lambda$  may be correlated  to properties of the microscopic particle, such as position, momentum and energy.
The set of projectors $\hat{P}_\lambda$ is exclusive ($\hat{P}_{\lambda} \hat{P}_{\lambda'} = 0, $ if $\lambda \neq \lambda'$). It is also exhaustive given that the event under consideration  has occurred; i.e., $\sum_\lambda \hat{P}_\lambda = \hat{P}$.

We also assume that  the system is initially ($t = 0$) prepared at a
state $|\psi_0 \rangle \in {\cal H}_+$, and that the time evolution is
governed by the self-adjoint Hamiltonian operator $\hat{H}$.

\subsection{Probability amplitudes with respect to the time of transition}
Quantum mechanical probabilities are defined in terms of squared amplitudes. Hence, in order to define probabilities for the time of transition, it is necessary to first construct the relevant amplitudes.
In particular, we employ the definitions of Sec. 2.1, in order to derive
 the probability amplitude $| \psi; \lambda, [t_1, t_2] \rangle$ that, given an initial state $|\psi_0\rangle$, a transition occurs at some instant in the time interval $[t_1, t_2]$ and a recorded value $\lambda$ is obtained for some observable.

 We first consider the case that the relevant time
interval is small, i.e., we set $t_1 = t$ and $t_2 = t + \delta t$, and we  keep only terms of leading  to $\delta t$.
Since the transition takes place within the interval $[t, t + \delta
t]$, at times prior to $t$ the state lay within ${\cal H}_-$. This is taken into account by evolving the initial state $|\psi_0 \rangle$ with the
restriction of the propagator into ${\cal H}_-$, that is, with the operator
\begin{eqnarray}
\hat{S}_t =  \lim_{N \rightarrow \infty}
(\hat{Q}e^{-i\hat{H} t/N} \hat{Q})^N. \label{restricted}
\end{eqnarray}

By assumption, the transition occurs at an instant within the time interval $[t, t+\delta t]$, after which a value $\lambda$ for a macroscopic observable is recorded. This means that in the time-interval $[t, t+\delta t]$ the amplitude transforms under the
full unitary operator for time evolution  $e^{-i \hat{H} \delta t}
\simeq 1 - i \delta t \hat{H}$. At time $t + \delta t$ the event
corresponding to $\hat{P}_{\lambda}$ is recorded, so the amplitude
is transformed by the action of $\hat{P}_{\lambda}$ (or of $\sqrt{\hat{P}_{\lambda}}$, if $\hat{P}_{\lambda}$ is not a
projector). For times greater than $t + \delta t$, there is no constraint, so the amplitude evolves
unitarily until some final moment $T$.

At the limit of small $\delta t$, the
successive operations above yield
\begin{eqnarray}
|\psi_0; \lambda, [t, t+ \delta t] \rangle =  - i \, \delta t \,
\,e^{-i\hat{H}(T - t)} \hat{P}_{\lambda} \hat{H} \hat{S}_t |\psi_0
\rangle. \label{amp1}
\end{eqnarray}

 The amplitude $|\psi_0; \lambda, [t, t + \delta t] \rangle$ is proportional to $\delta t$. Therefore, it defines  a {\em density} with respect to time: $|\psi_0;  \lambda, t \rangle := \lim_{\delta t \rightarrow 0}
\frac{1}{\delta t} | \psi_0; \lambda, [t, t + \delta t] \rangle$. From Eq. (\ref{amp1})
\begin{eqnarray}
|\psi_0;  \lambda, t \rangle = - i   \,e^{-i\hat{H}(T - t)}
\hat{P}_{\lambda} \hat{H} \hat{S}_t |\psi_0 \rangle = - i e^{- i
\hat{H} T} \hat{C}(\lambda, t) |\psi_0 \rangle, \label{amp2}
\end{eqnarray}
where the {\em class operator} $\hat{C}(\lambda, t)$ is defined as
\begin{eqnarray}
\hat{C}(\lambda, t) = e^{i \hat{H}t} \hat{P}_{\lambda} \hat{H}
\hat{S}_t. \label{class}
\end{eqnarray}

 Since the amplitude $|\psi_0;  \lambda, t \rangle $ is a density
with respect to the time of transition $t$, its integration with respect to $t$ is well-defined. Hence, the total amplitude that the transition occurred at {\em some time} in a time interval $[t_1,
t_2]$ is

\begin{eqnarray}
| \psi; \lambda, [t_1, t_2] \rangle = - i e^{- i \hat{H}T}
\int_{t_1}^{t_2} d t \hat{C}(\lambda, t) |\psi_0 \rangle.
\label{ampl5}
\end{eqnarray}

Eq. (\ref{ampl5}) involves the restricted propagator Eq. (\ref{restricted}), which may be difficult to compute in practice. However, there is an important regime where Eq. (\ref{ampl5}) simplifies significantly. We note that if $[\hat{P}, \hat{H}] = 0$, i.e., if the Hamiltonian
evolution preserves the subspaces ${\cal H}_{\pm}$, then $|\psi_0;
\lambda, t \rangle = 0$. For a Hamiltonian    of the form $\hat{H} = \hat{H_0} + \hat{H_I}$, where $[\hat{H}_0, \hat{P}] = 0$, and $H_I$ a perturbing interaction, we obtain to leading order in the perturbation
\begin{eqnarray}
\hat{C}(\lambda, t) = e^{i \hat{H}_0t} \hat{P}_{\lambda} \hat{H}_I
e^{-i \hat{H}_0t}, \label{perturbed}
\end{eqnarray}
and the restricted propagator Eq. (\ref{restricted}) does not appear in the amplitude Eq. (\ref{ampl5}).

\subsection{Probabilities with respect to the time of transition}

The amplitude  Eq. (\ref{amp2}) squared defines  the probability $p (\lambda, [t_1, t_2])\/$
that at some time in the interval $[t_1, t_2]$ a detection with
outcome $\lambda$ occurred
\begin{eqnarray}
P(\lambda, [t_1, t_2]) := \langle \psi; \lambda, [t_1, t_2] | \psi;
\lambda, [t_1, t_2] \rangle =   \int_{t_1}^{t_2} \,  dt
\, \int_{t_1}^{t_2} dt' \; Tr (e^{i\hat{H}( t - t')}
\hat{P}_{\lambda} \hat{H} \hat{S}^{\dagger}_t \hat{\rho}_0
\hat{S}_{t'} \hat{H} \hat{P}_{\lambda} ), \label{prob1}
\end{eqnarray}
where $\hat{\rho}_0 = |\psi_0\rangle \langle \psi_0|$.

However, the expression $P(\lambda, [t_1, t_2])$ does not correspond to a well-defined probability measure, because it
fails to satisfy the Kolmogorov additivity condition for probability measures. To see this, consider the probability corresponding to an
 interval $[t_1, t_3] = [t_1, t_2] \cup [t_2,
t_3]$. This equals
\begin{eqnarray}
P(\lambda, [t_1, t_3]) = P(\lambda, [t_1, t_2]) + P(\lambda, [t_2,
t_3])
+ 2 Re \left[ \int_{t_1}^{t_2} \,  dt \, \int_{t_2}^{t_3} dt'
Tr\left(\hat{C}(\lambda, t) \hat{\rho_0}\hat{C}^{\dagger}(\lambda,
t')\right)\right]. \label{add}
\end{eqnarray}
Hence, the Kolmogorov additivity condition $P(\lambda, [t_1, t_3]) = P(\lambda, [t_1, t_2]) + P(\lambda, [t_2,
t_3])$, necessary for a consistent definition of a probability measure, fails, unless

\begin{eqnarray}
2 Re \left[
\int_{t_1}^{t_2} \,  dt \, \int_{t_2}^{t_3} dt'
Tr\left(\hat{C}(\lambda, t) \hat{\rho_0}\hat{C}^{\dagger}(\lambda,
t')\right)\right] = 0 \label{decond}
 \end{eqnarray}
 In the consistent/decoherent histories framework, Eq. (\ref{decond}) is referred to as the
 {\em consistency condition} \cite{Omn1, Omn2, Gri}. It is the minimal condition necessary for the association of a consistent probability measure in histories. It appears naturally in the present framework, because we construct probabilities associated to properties of the system at different moments of time, that is, probabilities associated to histories.

   Eq. (\ref{decond}) does not hold for generic choices of $t_1, t_2$ and $t_3$.
   However, in a macroscopic system (or in a system with a macroscopic component) one expects that Eq. (\ref{decond}) holds with a good degree of approximation, given a sufficient degree of coarse-graining \cite{GeHa, hartlelo}. Thus, if the time of transition is associated to macroscopic records in a measurement apparatus, there exists a coarse-graining time-scale $\sigma$, such that the non-additive terms in Eq. (\ref{add}) are strongly suppressed if $ |t_2 - t_1| >> \sigma$ and $|t_3 - t_2| >> \sigma$. Then, Eq. (\ref{prob1}) does define a probability measure when restricted to intervals of size  larger than $\sigma$.

Hence, assuming a finite coarse-graining time-scale $\sigma$, such that Eq. (\ref{decond})is approximately valid for $ |t_2 - t_1| >> \sigma$ and $|t_3 - t_2| >> \sigma$, Eq. (\ref{prob1}) provides a consistent definition of a probability measure for the time of transition.

It is convenient to define the time-of-arrival probabilities by smearing the amplitudes
 Eq. (\ref{amp2}) at a time-scale of order $\sigma$ rather than employing probabilities for sharply defined time-intervals, as in Eq. (\ref{prob1}). Then, the time-of-transition  probabilities are expressed in terms of densities with respect to a continuous time variable.

To this end, we introduce a family of functions $f_{\sigma}(s)$,  localized around $s = 0$ with width $\sigma$, and normalized so that
$\lim_{\sigma \rightarrow 0} f_{\sigma}(s) = \delta(s)$. For example, one may employ the Gaussians
\begin{eqnarray}
f_{\sigma}(s) = \frac{1}{\sqrt{2 \pi \sigma^2}}
e^{-\frac{s^2}{2\sigma^2}}. \label{gauss}
\end{eqnarray}
The Gaussians Eq. (\ref{gauss}) satisfy the following equality.
\begin{eqnarray}
\sqrt{f_{\sigma}(t-s) f_{\sigma}(t-s')} = f_{\sigma}(t - \frac{s+s'}{2}) g_{\sigma}(s-s'), \label{eq2}
\end{eqnarray}
where
\begin{eqnarray}
g_{\sigma}(s) = \exp[-s^2/(8\sigma^2)]. \label{gsig}
 \end{eqnarray}

Using the functions $f_{\sigma}$, we  define the smeared amplitude $|\psi_0; \lambda, t\rangle_{\sigma}$ that is localized
 around the time $t$ with width $\sigma$, as
\begin{eqnarray}
|\psi_0; \lambda, t\rangle_{\sigma} := \int ds \sqrt{f_{\sigma}(s -t)}
|\psi_0; \lambda, s \rangle =  \int ds \sqrt{f_{\sigma}(s - t)} \hat{C}(\lambda, s) |\psi_0
\rangle, \label{smearing}
\end{eqnarray}
The square amplitudes
\begin{eqnarray}
P_{\sigma}(\lambda, t) = {}_{\sigma}\langle \psi_0; \lambda, t|\psi_0;
\lambda, t\rangle_{\sigma} = \int ds ds' \sqrt{f(s-t) f(s'-t)}
Tr \left[\hat{C}(\lambda, s)
 \hat{\rho}_0 \hat{C}^{\dagger}(\lambda, s')\right] \label{ampl6}
\end{eqnarray}
provide a well-defined probability measure: they are of the form
 $Tr[\hat{\rho}_0 \hat{\Pi}(\lambda, t)]$,
where
\begin{eqnarray}
\hat{\Pi}(\lambda, t) = \int ds ds' \sqrt{f_{\sigma}(s-t) f_{\sigma}(s'-t)} \hat{C}^{\dagger}(\lambda, s')
\hat{C}(\lambda, s) \label{povm2}
\end{eqnarray}
is a density with respect to both variables $\lambda$ and $t$.

The positive
operator
\begin{eqnarray}
\hat{\Pi}_{\tau}(N) = 1 - \int_0^{\infty} dt \int d \lambda
\hat{\Pi}_{\tau}(\lambda, t), \label{nodet}
\end{eqnarray}
 corresponds to the alternative $ N$ that no detection took place
in the time interval $[0, \infty)$. $\hat{\Pi}_{\tau}(N)$ together with the positive operators Eq. (\ref{povm2})
 define a Positive-Operator-Valued Measure (POVM). The POVM Eq. (\ref{povm2}) determines the probability density that a transition took place at time $t$, and that the outcome
$\lambda$ for the value of an observable has been recorded.

Using Eq. (\ref{eq2}), and setting $S = (s+s')/2$, $\tau = s -s'$, Eq. (\ref{ampl6} becomes
\begin{eqnarray}
P_{\sigma}( \lambda, t) = \int dS f_{\sigma}(t-S) \tilde{P}(\lambda, t), \label{smearing2}
\end{eqnarray}
where
\begin{eqnarray}
\tilde{P}(\lambda, t) \int d \tau g_{\sigma}(\tau) \left[\hat{C}(\lambda, t+\frac{\tau}{2})
 \hat{\rho}_0 \hat{C}^{\dagger}(\lambda, t- \frac{\tau}{2})\right] \label{pp}
\end{eqnarray}

Eq. (\ref{smearing2}) demonstrates that the probability distribution $P_{\sigma}$   is obtained by coarse-graining the classical probability distribution $\tilde{P}$ at a scale of $\sigma$.  For systems monitored at a time-scale much larger than $\sigma$ the two probability distributions essentially coincide. In that case, the probability density $\tilde{P}$ may be employed instead of $P_{\sigma}$. Moreover, if the resolution scale $\sigma$ is much larger than any timescale characterizing the microscopic system, we can take the limit $\sigma \rightarrow \infty$ in Eq. (\ref{pp}), i.e., we set $g_{\sigma} = 1$. The resulting probability distribution
\begin{eqnarray}
\tilde{P}(\lambda, t) = \int d\tau  Tr\left[\hat{C}(\lambda, t+\frac{\tau}{2})
 \hat{\rho}_0 \hat{C}^{\dagger}(\lambda, t- \frac{\tau}{2})\right] \label{pp2}
\end{eqnarray}
is independent of the coarse-graining scale $\sigma$.

Eq. (\ref{pp2}) is  a general expression for the time of transition in a quantum systems, which depends only on the initial state and the class operators $\hat{C}(\lambda, t)$, which are constructed solely from the Hamiltonian operator and the positive operators associated to the recorded observables. Thus, we have reduced the methodological problem of defining time-of-arrival probabilities to a problem of applying a general formula to experiments involving time-of-arrival measurements.

\subsection{Derivation of the time-of-arrival probability distribution}

Next, we employ Eq. (\ref{pp2}) for constructing probabilities associated to time-of-arrival measurements for a single particle. To this end, we select the relevant Hilbert space of the theory and the operators that appear in Eq. (\ref{pp2}), so that they describe the internal transitions of a  macroscopic apparatus that take place when a microscopic particle is detected.

 The system under consideration consists  of a microscopic particle and a macroscopic apparatus. The Hilbert space is the tensor product ${\cal F} \otimes {\cal H}_a$, where ${\cal H}_a$ is the Hilbert space describing the apparatus's degrees of freedom and ${\cal F}$  is the Fock space
\begin{eqnarray}
{\cal F} = {\bf C} \oplus {\cal H}_1 \oplus ({\cal H}_1  \otimes {\cal H}_1)_{S,A} \oplus \ldots \label{Fock}.
\end{eqnarray}
In Eq. (\ref{Fock}), ${\cal H}_1$ stands for the Hilbert space associated to  a single particle and $S$ and $A$ denote symmetrization and anti-symmetrization respectively. The reason we employ a Fock space is that in many detection processes, the microscopic particle is annihilated by the interactions at the detector. Hence, it is necessary to consider interactions where the number of microscopic particles is not conserved.

The Hamiltonian of the total system  is $\hat{H}_m \otimes 1 + 1 \otimes \hat{H}_{a} + \hat{H}_{int}$,
where $\hat{H}_m$ describes the dynamics of the
microscopic particle, $\hat{H}_{a}$ describes the dynamics of the apparatus and $\hat{H}_{int}$ is an interaction term.

  Finally, we  specify the macroscopic variables associated to particle detection. These correspond to degrees of freedom of the macroscopic apparatus and they are expressed in terms of the
 positive operators $1 \otimes \hat{\Pi}_{\lambda}$ on ${\cal F} \otimes {\cal H}_a$, labeled by the values $\lambda$ of a macroscopic observable. The operators  $\hat{\Pi}_{\lambda}$ are defined on
 ${\cal H}_{a}$ and they satisfy the completeness relation $\int d \lambda \hat{\Pi}_{\lambda} = \hat{P}$, where $\hat{P}$  is the projector onto the subspace ${\cal H}_+$.

A model for particle detection is defined through the specification of the Hilbert spaces ${\cal H}_1$ and ${\cal H}_m$ and of the operators $\hat{H}_m, \hat{H}_a$, $\hat{H}_{int}$ and $\hat{\Pi}_{\lambda}$. For each such specification, the POVM Eq. (\ref{pp2}) is uniquely constructed and it defines the time-of-arrival probabilities associated to this model. In this sense, the method we develop here is fully algorithmic, and it can be employed to construct different models according to the different physics of the detection scheme.

In particular, we assume that  the Hamiltonian $\hat{H}_m$ for the microscopic particle is invariant under spatial translations. Hence, it depends only on the particle's momentum operator.  We denote the particle's energy as a function of momentum (the particle's {\em dispersion relation}) as $\epsilon_{\bf p}$. We also consider an interaction Hamiltonian
\begin{eqnarray}
\hat{H}_{int} = \sum_i\int d^3 x \left[\hat{a}_i({\bf x}) \hat{J}^i_ +({\bf x}) + \hat{a}_i^{\dagger} ({\bf x}) \hat{J}^i_-({\bf x})\right], \label{hint}
\end{eqnarray}
where $\hat{a}_i({\bf x}), \hat{a}^{\dagger}_i({\bf x})$ are the annihilation and creation operators on ${\cal F}_s$,
$i$ labels non-translational degrees of freedom (e.g., spin), and  $\hat{J}^{\pm}({\bf x})$ are current operators
on ${\cal H}_{a}$ with support in the region $D$ where the detector is located. The interaction Hamiltonian $\hat{H}_{int}$ corresponds to a detection process where the microscopic particle is annihilated at the detector. This includes the cases that the particle is absorbed by an atom, or  that several product particles are created, or  a localized excitation is produced. Detection by absorption is not the only possibility, the incoming particle may be detected from energy or momentum transfer associated to a scattering process. In this case, the appropriate interaction Hamiltonian is
\begin{eqnarray}
\hat{H}_{int} = \sum_{ij} \int d^3 x \hat{\Lambda}_{ij}({\bf x}, {\bf x'} )  \hat{a}_i^{\dagger} ({\bf x'})\hat{a}_j({\bf x}), \label{hint4}
\end{eqnarray}
expressed in terms of a composite operator $\hat{\Lambda}_{ij}({\bf x}, {\bf x'} )$ defined on the Hilbert space ${\cal H}_a$. In this paper, we shall restrict our considerations to interaction Hamiltonians of the form (\ref{hint}), noting that the alternative Hamiltonian (\ref{hint4}) pose no particular problem for our approach.

 We place no restriction on the apparatus' Hamiltonian $\hat{H}_a$, except for the requirement that it does not lead to any spurious detection events, i.e. that detection records appear only when particles interact with the apparatus. This condition  that $[\hat{H}_{a}, \hat{P} ]   = 0$. It follows that $[1\otimes \hat{P}, \hat{H}_s \otimes 1 + 1 \otimes \hat{H}_{a}] = 0$; hence, the class operators $\hat{C}(\lambda, t)$ are obtained from  Eq. (\ref{perturbed}).

Finally, we consider a single-particle state for   the microscopic system
\begin{eqnarray}
|\psi_0\rangle = \sum_i \int d^3 x \psi_{0i}(x)\hat{a}^{\dagger}_i(x) |0\rangle \label{initial}
\end{eqnarray}
and an initial state of the apparatus $| \Psi_0\rangle$ stationary  with respect to the apparatus Hamiltonian $\hat{H}_a$. We assume that the state $| \Psi_0\rangle$
satisfies the condition $\hat{J}^-({\bf x}) |\Psi_0 \rangle = 0$, which guarantees that the only transitions in the detector are caused by the interaction with the microscopic particle.
 We also set the scale of energy so that $\hat{H}_{a}|\Psi_0 \rangle = 0$.

For the class of models specified above, the operator   Eq. (\ref{perturbed}) takes the form

\begin{eqnarray}
\hat{C}(\lambda, t) \left(|\psi_0 \otimes \rangle |\Psi_0\rangle \right)= \sum_i \int d^3 x \psi_i({\bf x}, t) |0 \rangle \otimes \left( e^{i \hat{H}_{a}t}\sqrt{\hat{\Pi}_{\lambda}} \hat{J}_i({\bf x}) |\Psi_0\rangle \right) ,
\end{eqnarray}
where $\psi({\bf x}, t)$ is the evolution, under $e^{- i \hat{H}_mt}$, of the state $|\psi_0\rangle$ in ${\cal H}_1$  in the position representation of a single-particle state.

Then,
\begin{eqnarray}
Tr \left(\hat{C}(\lambda, s) \hat{\rho}_0 \hat{C}^{\dagger}(\lambda, s') \right) &=& \sum_{ij}\!\!\int \! \! d^3 x d^3 x' \psi_i({\bf x}, s) \psi^*_j({\bf x'}, s')
\nonumber \\
&\times& \langle \Psi_0| \hat{J}_j^-({\bf x'}) \sqrt{\hat{\Pi}_{\lambda}} e^{- i \hat{H}_{a}(s' - s)} \sqrt{\hat{\Pi}_{\lambda}} \hat{J}_i^+({\bf x})|\Psi_0\rangle. \label{crc}
\end{eqnarray}

In order to simplify the analysis, we ignore the non-translational degrees of freedom of the particle and we reduce the system to one spatial dimension, which corresponds to the axis connecting the source and the detection region. Substituting Eq. (\ref{crc}) in Eq. (\ref{pp}),  the probability density takes the form
\begin{eqnarray}
\tilde{P}(\lambda, t) = \langle \psi_t|\hat{S}(\lambda) |\psi_t \rangle \label{plt}
\end{eqnarray}
where $\hat{S}(\lambda)$ is an operator defined in terms of its matrix elements in the momentum basis
\begin{eqnarray}
 \langle p' |\hat{S}(\lambda)|p \rangle = \int d\tau  e^{ -i (\epsilon_p+\epsilon_{p'})\tau/2} \int dx dx' e^{i px -i p'x'} \nonumber \\
\times \langle \Psi_0| \hat{J}^-({\bf x'}) \sqrt{\hat{\Pi}_{\lambda}} e^{ i \hat{H}_{a}\tau} \sqrt{\hat{\Pi}_{\lambda}} \hat{J}^+({\bf x})|\Psi_0\rangle. \label{fpp}
\end{eqnarray}

In a time-of-arrival measurement, the pointer variable $\lambda$ corresponds to position $X$ along the particle's axis of motion.
We assume the detector is located at a macroscopic distance $L$ from the source, and that the accuracy of the detector's position sampling is of order $\delta$. Then, we consider positive operators $\sqrt{\hat{\Pi}_L}$ corresponding to an unsharp Gaussian sampling of position at $X = L$,

\begin{eqnarray}
\sqrt{\hat{\Pi}_L} = \frac{1}{(\pi \delta^2)^{1/4}}\sum_a \int dX e^{-\frac{(L-X)^2}{2 \delta^2}} |X, a\rangle \langle X,a |, \label{po}
\end{eqnarray}

 The index $a$ in Eq. (\ref{po}) refers to the degrees of freedom of the apparatus other than the pointer variable. Substituting Eq. (\ref{po}) into Eq. (\ref{fpp}), we obtain the general form for the time-of-arrival distribution for the particle.

We note that the familiar probability current Eq. (\ref{current0}) for a non-relativistic particle of mass $m$ is of the form Eq. (\ref{plt}), with $\hat{S}(L) = \frac{1}{2} \left[ \hat{p} \delta(\hat{x} - L) + \delta(\hat{x} - L) \hat{p} \right]$.  However, the operator $\hat{S}(L)$ of Eq. (\ref{current0}) is non-local when acting on configuration-space wave-functions, making the probabilities (\ref{plt}) {\em non-local} functionals of the wave-functions $\psi_t(x)$. Thus, an interpretation of $\hat{S}(L)$ as a probability-current operator is untenable.

Eqs. (\ref{plt}) and (\ref{fpp}) define a probability density for the time of arrival that applies to any detection scheme, in which the particle is annihilated. The explicit form of the probability density above depends on the physics of particle detection. In the following section, we will consider specific models for the detector and its interaction with the microscopic particle.

\section{Time-of-arrival probabilities for different models of particle detection}

In this section, we explain how Eq. (\ref{plt}) is to be employed for the derivation of explicit time-of-arrival probabilities associated to specific experiments. We derive such probabilities for different detector models, and we demonstrate that there exists a special regime in which all information about the detector is encoded in a single function of momentum, the absorption coefficient $\alpha(p)$. From these models, we obtain a generalization of Kijowski's probability distribution, valid for any dispersion relation for the microscopic particle. Finally, we adapt our formalism for the description of the time-of-arrival in high-energy physics, where the detection process involves the creation of several product particles on the detector.

\subsection{Three particle-detection models}

The time-of-arrival probability distribution Eq. (\ref{plt}) does not depend only on characteristics of the particle (the initial state and the dispersion relation), but also on characteristics of the detector. The latter are incorporated into  (i) the Hamiltonian $\hat{H}$  that describes the detector's self-dynamics, (ii) the current operator $\hat{J}_+({\bf x})$ that describes the interaction of the microscopic particle with the detector, (iii) the initial quantum state of the detector $|\Psi_0\rangle$, and (iv) the set of positive operators $\hat{\Pi}_{\lambda}$ that correspond to the  pointer variables of the apparatus. Hence, the description of any particular time-of-arrival experiment involves a modeling of the detector through a specification of the aforementioned mathematical objects. In what follows, we will consider three different types of detector models, and study the properties of the associated time-of-arrival probabilities.

\subsubsection{Detection of a coherent particle excitation}

First, we consider the case that the detection of a microscopic particle is accompanied by the creation of a particle-like excitation at the detector. The relevant pointer variable is the excitation's position $X$ at the locus of the interaction. For example, the excitation may correspond to an excited nucleus or an atom. We assume that the pointer-variable $X$ is approximately an autonomous variable, in the sense the interaction with the remaining degrees of freedom at the apparatus is weak. Thus, we can identify the subspace ${\cal H}_{a+}$ of ${\cal H}_a$ of accessible states in the apparatus after a particle detection with the Hilbert space $L^2(R,dX)$ of a single particle.

{\em The Hamiltonian.} Assuming negligible interaction with the remaining degrees of freedom, the pointer variable evolves unitarily. We consider a Hamiltonian on  ${\cal H}_{a+}$  that describes a non-relativistic particle of effective mass $\mu_+$,

 \begin{eqnarray}
 \hat{H}_a = E_0 + \frac{1}{2\mu_*} \hat{P}^2, \label{hm}
  \end{eqnarray}
  where $\hat{P}$ is the excitation's momentum and $E_0$ is a constant corresponding to the energy gap in the detector, that is the energy required for the creation of the excitation.

{\em Initial state and current operator.} Since the pointer variable $X$ is correlated to the point $x$ where the microscopic particle has been annihilated, we expect that $\langle X|\hat{J}^+(x)|\Psi_0\rangle \simeq 0$, if $|x - X|$ is significantly larger than $\delta$. We, therefore, write $\langle X|\hat{J}^+(x)|\Psi_0\rangle = u(x-X)$, where $u(x)$  is a function that vanishes  for $|x| >> \delta$. The function $u(x)$ incorporates all information about the initial state of the apparatus and its interaction with the microscopic particle. Later, we shall establish that $u(x)$ is closely related to an absorption coefficient associated to the detector.

{\em Pointer variables.} For the Hilbert space $L^2(R,dX)$, the positive operators (\ref{po}) simplify
\begin{eqnarray}
\sqrt{\hat{\Pi}_L} = \frac{1}{(\pi \delta^2)^{1/4}} \int dX e^{-\frac{(L-X)^2}{2 \delta^2}} |X \rangle \langle X |. \label{po2}
\end{eqnarray}

With the assumptions above,
 Eq. (\ref{fpp}) becomes
\begin{eqnarray}
 \langle p'|\hat{S}(L)|p\rangle = \frac{1}{\sqrt{\pi \delta^2}}\int d \tau e^{ -i (\epsilon_p+\epsilon_{p'})\tau/2} \int dx dx' \int dX dX' e^{-\frac{(X-L)^2 +(X'-L)^2}{2 \delta^2}} \nonumber \\
 \times u(x-X) u^*(x'-X')
 e^{ipx -ip'x'}  G(X-X', \tau), \label{fpp2}
\end{eqnarray}
where
\begin{eqnarray}
G(X-X', \tau) := \langle X|e^{ i (E_0 + \frac{\hat{P}^2}{2 \mu_*})}|X'\rangle = \sqrt{\frac{\mu_*}{-2\pi i \tau}} e^{-i \frac{\mu_*(X-X')^2}{2\tau} + i E_0 \tau }.
\end{eqnarray}
Carrying out the integrations over $x, x'$ and $(X+X')/2$ in Eq. (\ref{fpp2}), we obtain
\begin{eqnarray}
 \langle p'|\hat{S}(L)|p\rangle =  e^{-\delta^2(p-p')^2/4}\tilde{u}(p) \tilde{u}^*(p') e^{i(pL -ip'L}\nonumber \\
\times \int d\tau  e^{-i(\epsilon_p + \epsilon_{p'})\tau/2} \int dZ e^{-\frac{Z^2}{4\delta^2}} e^{i(p+p')Z/2} G(Z, \tau), \label{fpp3}
\end{eqnarray}
where $Z = X - X'$ and $\tilde{u}(p)$ is the Fourier transform of $u(x)$.

We consider initial states $|\psi_0\rangle$ emitted from the source at $x = 0$ with support on positive values of $p$. For macroscopic values of $L$,  the probability density Eq. (\ref{plt})  is strongly suppressed if $t < 0$. Thus, when we consider the time-integrated time-of-arrival probability, we can extend the range of integration from $[0, \infty)$ to $(-\infty, \infty)$. We then obtain

\begin{eqnarray}
\tilde{P}(L) := \int_0^{\infty} dt \tilde{P}(t,L)\simeq \int_{-\infty}^{\infty} dt \tilde{P}(t,L)  = \int \frac{dp}{2 \pi} \frac{\langle p| \hat{S}(L)|p\rangle}{|v_p|} |\tilde{\psi}_0(p)|^2, \label{intp}
\end{eqnarray}
i.e., the time-integrated probability   is $L$-independent. In Eq. (\ref{intp}), $\tilde{\psi}_0$ is the particle's initial state in the momentum representation, and $v_p = \partial \epsilon_p/\partial p$ is the particle's velocity.

The time-integrated probability $\tilde{P}(L)$ is a density with respect to $L$, hence, for a detector of  size $d$ in the $x$-direction, the total fraction of detected particles equals $\tilde{P}(L) d$. For a monochromatic initial state with momentum $p_0$, $|\tilde{\psi}_0|^2 \simeq 2 \pi \delta (p - p_0)$, the total detection probability equals $d \langle p_0| \hat{S}(L)|p_0\rangle/|v_{p_0}|$. This implies that we can define an {\em absorption coefficient } $\alpha(p)$ for the detector\footnote{The definition of an absorption coefficient depends crucially on the property, established earlier, that the time-integrated probability is $L$-independent. }defined, standardly, as the fraction of incoming particles with momentum $p > 0$ absorbed per unit length of the absorbing medium

\begin{eqnarray}
\alpha(p) = \frac{\langle p| \hat{S}(L)|p\rangle}{|v_p|}. \label{absdef}
\end{eqnarray}

Next, we evaluate the matrix elements of the operator $\hat{S}(L)$, Eq. (\ref{fpp3})   at the limit   $E \mu_* \delta^2 << 1$, that is, under the assumption of a very narrow localization of the particle excitation. In this regime, the integral over $Z$ in Eq. (\ref{fpp3}) is approximated by
\begin{eqnarray}
\int dZ e^{-\frac{Z^2}{4\delta^2}} e^{i\frac{p+p'}{2}Z} G(Z, \tau)  \simeq \sqrt{\frac{\mu_*\delta^2}{-i \tau}} \exp\left[ -\frac{\delta^2}{4} (p+p')^2 + i E_0 \tau \right].
\end{eqnarray}

Integrate  Eq. (\ref{fpp3}) over $\tau$, we obtain
\begin{eqnarray}
 \langle p'|\hat{S}(L)|p\rangle = \frac{ \sqrt{\pi \mu_* \delta^2}}{\sqrt{\frac{\epsilon_p+\epsilon_{p'}}{2} - E_0}} \tilde{u}(p) \tilde{u}^*(p') e^{-\frac{\delta^2}{2}(p^2+p^{\prime 2})} e^{ipL-ip'L}. \label{fpp4}
\end{eqnarray}

 Eq. ({\ref{fpp4}) simplifies
for initial  states $\tilde{\psi_0}(p)$ with mean momentum $p_0$ such that $E_0 << \epsilon_{p_0}$ and with momentum spread $\Delta p$ such that  $\Delta p << p_0$. In this regime, $\epsilon_p + \epsilon_{p'} \simeq 2 \sqrt{\epsilon_p \epsilon_{p'}}$. The probability density Eq. (\ref{plt}) becomes
\begin{eqnarray}
\tilde{P}(L,t) = K \left|\int \frac{dp}{2\pi} e^{-\delta^2 p^2/2}\frac{\tilde{u}(p)}{\epsilon_p^{1/4}} \tilde{\psi}_0(p) e^{ipL-i\epsilon_pt}\right|^2 \label{plt2}
\end{eqnarray}
where $K =  \sqrt{\pi \delta^2\mu_*/2}$.

Eq. (\ref{plt2}) is straightforwardly generalized to initial mixed states of the form
\begin{eqnarray}
\hat{\rho} = \int dp_0 f(p_0)| \psi_{p_0} \rangle \langle \psi_{p_0}|, \label{rhho}
\end{eqnarray}
 where $|\psi_{p_0}\rangle $ denotes an overcomplete set of pure states (for example, coherent states) with mean momentum $p_0$ and spread $\Delta p$, such that $\Delta p << p_0$, and $f(p_f0)$ is a positive-valued function.  To see this, let as denote the probability density Eq. (\ref{plt2}) evaluated for initial state $|\psi_{p0} \rangle$ as $\tilde{P}(L,t;\psi_{p0})$. Then the probability density $\tilde{P}(L,t;\hat{\rho})$ associated to the initial state $\hat{\rho}$, Eq. (\ref{rhho}) is
\begin{eqnarray}
\tilde{P}(L,t;\hat{\rho}_0) = \int dp_0 f(p_0) \tilde{P}(L,t;\psi_{p0}). \label{plt2c}
\end{eqnarray}
In particular, there is no requirement that the momentum spread of the initial mixed state $\hat{\rho}$ be smaller than its mean momentum for Eq. (\ref{plt2b}) to hold.

The presence of the cut-off factor $e^{-\delta^2 p^2}$ in Eq. (\ref{plt2}) implies that the detection of particles with momenta $p >> \delta^{-1}$ is strongly suppressed. For values of $p<<\delta^{-1}$, such that the suppression factor $e^{-\delta^2 p^2}$ can be ignored, the current operator $\hat{S}(L)$ becomes
\begin{eqnarray}
\hat{S}(L) = K \tilde{u}^*(\hat{p}) \hat{h}^{-1/4} \delta (\hat{x} - L) \hat{h}^{-1/4} \tilde{u}(\hat{p}),
\end{eqnarray}
where $\hat{h}$ is the particle's Hamiltonian.

The absorbtion coefficient $\alpha(p)$   corresponding to Eq. (\ref{plt2} is
\begin{eqnarray}
\alpha(p) = K \frac{|\tilde{u}(p)|^2}{|v_p|\sqrt{\epsilon_p}}, \label{abso}
\end{eqnarray}
where $v_p = \partial \epsilon_p/\partial p$ is the particle's velocity.

Eq. (\ref{abso}) implies that
\begin{eqnarray}
\tilde{u}(p) = \sqrt{\frac{\alpha(p) |v_p| \sqrt{\epsilon_p}}{K}} e^{i \theta(p)},
 \end{eqnarray}
 where $\theta(p)$ is a momentum-dependent phase. The phase $\theta(p)$ contributes to the total phase factor $i[\theta(p) + pL - \epsilon_p t]$ of the integral in Eq. (\ref{plt2}). The variation of $\theta(p)$ is expected to define a microscopic time-scale much smaller than $L$ ($|\theta'(p)| << L$), hence, its contribution to the phase of Eq. (\ref{plt2}) is negligible. Then, Eq. (\ref{plt2}) becomes
\begin{eqnarray}
\tilde{P}(L,t) = \left| \int \frac{dp}{2\pi} \sqrt{\alpha(p) |v_p|} \tilde{\psi}_0(p) e^{ipL-i\epsilon_pt}\right|^2. \label{plt2b}
\end{eqnarray}
Thus, in this regime all information about the measurement apparatus is contained in the absorption coefficient $\alpha(p)$.

\subsubsection{Detection of a decoherent excitation}

 A more realistic description of detection through the creation of a particle-like excitation involves taking into account the influence of the remaining degrees of freedom of the apparatus in the evolution of the pointer variable. To this end, we treat the other degrees of freedom as an environment which results to non-unitary dynamics on the excitation. The evolution of the pointer variable $X$ is then treated using the theory of quantum open systems.

 The model we consider here involves the same Hilbert space, and the same expressions for the initial state, the interaction current and the positive operators $\hat{\Pi}_L$ as the model of Sec. 3.1.1. The difference is in the choice of the Hamiltonian. We model the effect of the environment by introducing a stochastic term in the  Hamiltonian $\hat{H}_a$ of Eq. (\ref{hm}) for the pointer variable $\hat{X}$,
\begin{eqnarray}
\hat{H}_a = E_0 + \frac{1}{2\mu_*} \hat{P}^2 + \hat{P} \xi(t),
\end{eqnarray}

where  $\xi(t)$ is a Markovian process satisfying
\begin{eqnarray}
{\cal M} [\xi(t)] = 0, \hspace{1cm}
{\cal M} [ \xi(t) \xi(t')] = D \delta(t-t'). \label{sto}
\end{eqnarray}
In Eq. (\ref{sto}), ${\cal M}$ denotes ensemble average and $D$ is a phenomenological constant.

In this model, the matrix elements of the operator $\hat{S}(L)$ are of the form Eq. (\ref{fpp2}), where
\begin{eqnarray}
G(X-X', \tau) := \langle X'|{\cal M} [\hat{U}(s') \hat{U}^{\dagger}(s)]|X \rangle,
\end{eqnarray}
where $\tau = s' - s$.

 We evaluate the kernel $G$ to second-order in perturbation theory for the noise:
\begin{eqnarray}
G(X-X', \tau) = \langle X'|e^{i\Delta \tau + i \frac{\hat{P}^2}{2\mu_*} \tau - \frac{D}{2\mu^2} \hat{P}^2 |\tau|}|X \rangle = \sqrt{\frac{\mu_*}{-2\pi i (\tau+iD |\tau|/\mu_*)}} e^{-i \frac{\mu_*(X-X')^2}{2(\tau+iD |\tau|/\mu_*)} + i E_0 \tau }.
\end{eqnarray}

The regimes $D/\mu_* << 1$ of weak coupling to the environment corresponds to leading order in $D/\mu_*$ to the model studied in Sec. 3.1.1. Here, we consider the opposite limit  $D/\mu_*>>1$ of strong coupling to the environment. In this regime,   the exponential damping behavior dominates  in $G(X-X', \tau)$, and the excitation loses rapidly all quantum coherence. In this regime,
\begin{eqnarray}
G(X-X', \tau) \simeq \frac{\mu_*}{\sqrt{2 \pi D |\tau}|} e^{ - \frac{\mu_*^2 (X-X')^2}{2 D |\tau|}+i E_0 \tau}.
\end{eqnarray}
Thus,
\begin{eqnarray}
\int dZ e^{-\frac{Z^2}{4\delta^2}+i (p+p')Z/2} G(Z, \tau) = (1+ \frac{D|\tau|}{2 \delta^2 \mu_*^2})^{-1/2} \exp\left[- \frac{ \delta^2 (p+p')^2}{4(1+ \frac{2 \mu_*^2 \delta^2}{D|\tau|})}\right]. \label{zinter}
\end{eqnarray}
In the integration of the right-hand side of Eq. (\ref{zinter}) over $\tau$, the dominant contribution comes from values of $|\tau| < < \tau_{dec} :=  \mu_*^2 \delta^2/D$. The time scale $\tau_{dec}$ is a decoherence time scale: virtual processes with a difference $\Delta t$ in the time of arrival, do not contribute to the total probability if $\Delta t >> \tau_{dec}$.

For sufficiently strong coupling $D$ to the environment, the damping term in Eq. (\ref{zinter}) dominates and
 the right-hand-side term of Eq. (\ref{zinter}) equals $\exp \left[- \frac{D(p+p')^2}{2 \mu_*^2}|\tau|\right]$. Integration over $\tau$ in Eq. (\ref{fpp2}) leads to a multiplicative term $\frac{4 \mu_*^2}{D(p+p')^2}$, and
  Eq. (\ref{fpp2})   becomes

\begin{eqnarray}
\langle p'|\hat{S}(L)|p\rangle = \frac{4 \mu_*^2}{D(p+p')^2} \tilde{u}(p) \tilde{u}^*(p') e^{-\frac{\delta^2(p-p')^2}{4}} e^{ipL- ip'L}. \label{fpp6}
\end{eqnarray}
In Eq. (\ref{fpp6}), contributions from different momenta are suppressed by a factor $e^{-\delta^2(p-p')^2/4}$. This is the most significant qualitative difference from the model of a coherent excitation of Sec. 3.1.1, in which the coherence of momentum superpositions is preserved.

Since the diagonal matrix elements of $\hat{S}(L)$ are independent of $L$, the consideration of initial states with support on positive values of $p$, leads to a definition of the absorption coefficient
\begin{eqnarray}
\alpha(p) = \frac{\mu_*^2}{D v_p p^2} |\tilde{u}(p)|^2. \label{ab2}
\end{eqnarray}

For pure states with mean momentum $p_0$  and  momentum spread $\Delta p$, such that $\Delta p << \delta^{-1}$ and $\Delta p << p_0$, Eq. (\ref{fpp6}) leads again to Eq. (\ref{plt2b}), with $\alpha(p)$ given by Eq. (\ref{ab2}), which is valid for all   mixed states of the form Eq. (\ref{rhho}).

\subsubsection{Particle detection from energy absorption}
The third model to be considered here describes an apparatus where the microscopic particle is detected through the excitation of  the detector's energy levels. An variation of this model has been studied in  detail in Ref. \cite{AnSav11}, where it was shown that it corresponds to a Glauber-type photo-detector in one regime and to a macroscopic Unruh-Dewitt detector in another. The main features of this model are the following.

{\em Hamiltonian.} This model requires no specific form of the Hilbert space ${\cal H}_a$ of the apparatus degrees of freedom and of the corresponding Hamiltonian $\hat{H}_a$. We only assume that the energy eigenstates spanning ${\cal H}_a$ correspond to a density of states $w(E)$.

{\em Pointer variable.} We consider a single pointer variable $\hat{X}$ corresponding to the location of the detector, and we assume that $\hat{X}$ is invariant under time translations generated by the Hamiltonian $\hat{X}$: $[\hat{X}, \hat{H}] = 0$. For example. $\hat{X}$ may describe different independent sub-detectors, whose location is fixed in space. We  employ Eq. (\ref{po}) for the positive operators corresponding to position sampling.

{\em Initial state and current operator.} Denoting the eigenstates of the Hamiltonian $\hat{H}_a$ by $|a \rangle$, Eq. (\ref{fpp}) for the time-of-arrival probabilities involves matrix elements of the form $\langle X, a|\hat{J}_+(x)|\Psi_0\rangle$. We assume that these matrix elements depend on the label $a$ only through the energy eigenvalue $E(a)$ associated to $a$, that is we express
\begin{eqnarray}
\langle X, a|\hat{J}_+(x)|\Psi_0\rangle = u(x-X, E(a)),
\end{eqnarray}

for some function $u(x, E)$ of position $x$ and energy$E$.

 With the above assumptions, Eq. (\ref{fpp} becomes

\begin{eqnarray}
 \langle p'|\hat{S}(L)|p\rangle = \sqrt{2} \pi  e^{-\frac{\delta^2(p-p')^2}{4}} \int dE w(E) \tilde{u}_{\delta}(p,E) \tilde{u}^*(p',E) e^{ipL-ip'L} \delta(E - \frac{\epsilon_p+\epsilon_{p'}}{2}). \label{fpp5}
\end{eqnarray}

The diagonal matrix elements of $\hat{S}(L)$ are independent of $L$, leading to the following expression for the absorption coefficient
\begin{eqnarray}
   \alpha(p) = \sqrt{2} \pi  \frac{w(\epsilon_p) |\tilde{u}_{\delta}(p,\epsilon_p)|^2}{|v_p|}. \label{ab3}
\end{eqnarray}
Again, for pure states with mean momentum $\tilde{p}$  and  momentum spread $\Delta p$, such that $\Delta p << \delta^{-1}$ and $\Delta p << \bar{p}$, Eq. (\ref{fpp5}) leads  to Eq. (\ref{plt2b}), with $\alpha(p)$ given by Eq. (\ref{ab3}).

\subsection{General expressions for the time-of-arrival probability}

As demonstrated in the three models of Sec. 3.1, the probability densities for the time of arrival depend sensitively on the detailed physics of the particle detection scheme. In what follows, we shall show that, nonetheless, there exist particular regimes where they take similar form, and that one particular regime corresponds to a generalization of Kijowski's probability distribution \cite{Kij}.

\subsubsection{Initial states with sharp momentum.}
    For initial states $\tilde{\psi}_0(p)$ with mean momentum $p_0$ and spread $\Delta p$  and $\Delta p << p_0$,  Eqs. (\ref{fpp4}), (\ref{fpp5}) and (\ref{fpp6}) define  a probability density of the form
\begin{eqnarray}
\tilde{P}(L,t) \simeq K(p_0) \left|\int dp  \tilde{\psi}_0(p) e^{ipL-i\epsilon_pt}\right|^2 + O(\Delta p /\bar{p}), \label{plt4}
\end{eqnarray}
where   $K(p_0)$ depends on the special characteristics of each measurement scheme, and it is proportional to the absorption coefficient $\alpha(p_0)$. The total detection probability depends on the measurement scheme, however, the time-of-arrival probabilities conditioned upon detection depend only on the initial state. For initial states $\psi_0$ centered around $x = 0$, the probability density Eq. (\ref{plt4}) is peaked around the stationary phase point $L - (\partial \epsilon_p/\partial p) t$, and corresponds to a mean time of arrival $t_{cl} = L/v_p$.

\subsubsection{The classical regime.}

 Before studying the classical limit of the time-of-arrival probabilities derived in Sec. 3.1, we first examine a version of the classical time-of-arrival problem.  We consider an ensemble of classical free particles, described at $t = 0$ by a phase space probability  distribution $\rho_0(x,p)$, where $x$ and $p$ are a particle's position and momentum, respectively. The time of arrival at $x = L$ is an observable on the state space, defined as $ t = (L - X)/v_p$. Thus, the probability density for the time of arrival is given by
\begin{eqnarray}
P(t) = \int dx dp \delta(t - \frac{L - x}{v_p}) \rho_0(x,p). \label{pcl}
\end{eqnarray}

The corresponding equation for quantum mechanical particles is obtained from Eq. (\ref{plt}) by expressing the initial state $\psi_0$ in terms of its Wigner function $W_0(x,p)$,
\begin{eqnarray}
\tilde{P}(L, t) = \int dx dp W_{S}(t, x, p) W_0(x,p),
\end{eqnarray}
where
\begin{eqnarray}
W_S(t, x, p) = \int \frac{d \xi}{2\pi} e^{ix\xi} \langle p - \frac{\xi}{2}|e^{i\hat{H}_m t}\hat{S}(L) e^{-i\hat{H}_m t}|p + \frac{\xi}{2} \rangle \label{ws}
\end{eqnarray}
is the Wigner transform of the operator $e^{i\hat{H}_m t}\hat{S}(L) e^{-i\hat{H}_m t}$.

We compute Eq. (\ref{ws}) the  operator $\hat{S}(L)$ of  Eq. (\ref{plt2b}) which, as we saw, applies for a large class of initial states in different detection schemes. The integral Eq. (\ref{ws}) contains term of the form $v_{p\pm\frac{\xi}{2}}, \alpha(p \pm \frac{\xi}{2})$ and $\epsilon_{p\pm \frac{\xi}{2}}$. Expanding those terms around $p$ and keeping the leading order terms with respect to $\xi$, we obtain

\begin{eqnarray}
W_S(t, X, P) =   \alpha (p) \delta(t - \frac{L - x}{v_p}). \label{wf2}
\end{eqnarray}
Hence, the time of arrival distribution becomes
\begin{eqnarray}
\tilde{P}(t,L) = \int dx dp  \alpha(p) \delta(t - \frac{L - x}{v_p}) W_0(x,p). \label{wf3}
\end{eqnarray}
Eq. (\ref{wf3}) corresponds to the classical probability distribution (\ref{pcl}) modified by the absorption coefficient $\alpha(p)$ that takes into account the momentum dependence of the particle's detection probability. Eq. (\ref{wf3}) is not normalized to unity; to normalize, one has to divide by the total probability of detection $\int dt \tilde{P}(t,L)$.

We also calculate the first quantum correction by keeping terms proportional to $\xi^2$ in the integral Eq. (\ref{ws}). We obtain

\begin{eqnarray}
W_S(t, X, P) =   \alpha (p) \delta(t - \frac{L - x}{v_p}) - \frac{\alpha \alpha'' -  (\alpha')^2 + v_p v_p'' - (v'_p)^2 }{8\alpha v_p^3} \partial_t^2 \delta(t - \frac{L - x}{v_p}) + \ldots,
\end{eqnarray}
where the primes denote differentiation with respect to the momentum $p$.

\subsubsection{Generalization of Kijowski's POVM}

The total integral over time of the probability densities for the time of arrival corresponds to the total number of particles detected from a detector at $x = L$. In general, the total probability depends on the properties of the initial state.
 This implies that the normalized probability density $\tilde{P}(L, t)/\int dt \tilde{P}(L, t)$ is {\em not} a linear functional of the initial state. This is to be expected, since $\tilde{P}(L,t)/\int dt \tilde{P}(L,t)$ corresponds to the probability distribution for the time-of-arrival {\em conditioned} upon the particle having been detected.

 There is, however, a special regime where the probability density for the time of arrival can be normalized to unity by dividing with a constant. This regime corresponds to the domain of validity of  Eq. (\ref{plt2b})(for all three models of Sec. 3.1),  in the special case that the absorption coefficient is a constant. In this case, we obtain the probability density
\begin{eqnarray}
\tilde{P}(L,t) = \left| \int \frac{dp}{2 \pi} \sqrt{|v_p|} \tilde{\psi}_0(p) e^{ipL-i\epsilon_pt}\right|^2, \label{plt3}
\end{eqnarray}
which is normalized to unity for all initial states with support on positive values of momentum.

For a non-relativistic particle ($\epsilon_p = \frac{p^2}{2m}$), Eq. (\ref{plt3}) coincides with Kijowski's POVM for the time-of-arrival. Thus, Kijowski's POVM is identified as an ideal time-of-arrival POVM, in which the probability of detection does not depend on the incoming particles' momentum.
This interpretation is consistent with the results of Ref. \cite{AnSav06}, where Kijowski's POVM was obtained by modeling the detector as a totally absorbing surface for particles of {\em all} momenta, defined by Dirichlet boundary conditions at $x = L$.

  Eq. (\ref{plt3}) provides the generalization of Kijowski's POVM, valid for a general dispersion relation. The corresponding current operator is
\begin{eqnarray}
\hat{S}(L) = \sqrt{|\hat{v}|} \delta(\hat{x} - L) \sqrt{|\hat{v}|},
\end{eqnarray}
where $\hat{v} = (\partial \epsilon/\partial p)(\hat{p})$ is the velocity operator.

For a relativistic particle of mass $m$, $\hat{v} = \hat{p} (\hat{p}^2+m^2)^{-1/2}$; the probability density Eq. (\ref{plt3}) is not invariant under Lorentz transformations. This is to be expected, since the calculation explicitly involves the rest-frame of the detector, and the initial state of the detector's degrees of freedom is not Lorentz covariant.

\subsection{Time-of-arrival probabilities in high-energy particle reactions}

The simple models studied in Sec. 3.1 allowed us to derive explicit expressions for the time-of-arrival probability and led to the generalization of Kijowski's probability distribution. Here, we expand the scope of our approach, in order to define time-of-arrival probabilities for experiments in which the particle under consideration is detected through high-energy particle reactions. Our aim is to
\begin{enumerate}
\item  Demonstrate that our method applies to high-energy processes and that it can incorporate quantum-field theoretic interactions.
\item Provide a measurement-theoretic characterization of the particle detection process, which, in principle, can lead to explicit mathematical modeling of realistic particle detectors.

\item Demonstrate that Eq. (\ref{plt2b}) for the time-of-arrival probabilities is of general validity, and not restricted to the models of Sec. 3.1.
\end{enumerate}

To this, we consider a measurement scheme where particles, denoted as $A$ are produced from a source around $x = 0$ and propagate towards a detector at distance $L$ from the source, where they are detected by means of the process

\begin{eqnarray}
A + B_1 + \ldots B_M \rightarrow D_1  \ldots + D_N. \label{proc}
\end{eqnarray}
where $B_m, D_n$ are particles (different from the $A$ particles),
labeled by the indices $m = 1, \ldots, M$ and $n = 1, \ldots, N$.   The interaction of the $A$ particles with the $B_m$ particles produces the   particles $D_n$, which are the ones that are being detected. Relevant observables are the $D_n$-particles' time of detection, position, momentum and so on. These observables are determined through macroscopic pointer variables in the detector.

\medskip

 The Hilbert space ${\cal H}_{tot}$ associated to the process Eq. (\ref{proc}) are described  is  as a tensor product
 ${\cal H}_{tot} = {\cal
H}_A \otimes {\cal H}_{r}$. ${\cal H}_A$ is the Fock space ${\cal
F}({\cal H}_{1})$ corresponding to the $A$-particles. Explicitly, we write
\begin{eqnarray}
{\cal F}({\cal H}_{1}) = {\bf C} \oplus H_{1} \oplus
(H_{1}\otimes H_{1})_{S, A} \oplus (H_{1}\otimes H_{1} \otimes
H_{1})_{S, A} \oplus \ldots,
\end{eqnarray}
where $S$ refers to symmetrization (bosons) and $A$ to
antisymmetrization (fermions) and $H_{1A}$ is the Hilbert space describing a
single $A$ particle.
The degrees of freedom of the $B_m $ and $D_n$ particles are incorporated into the Hilbert space ${\cal H}_r$. In general, ${\cal H}_r$ is a tensor product of Fock spaces, one for each field other than $A$, participating in the process Eq.  (\ref{proc}) and of a Hilbert space describing other degrees of freedom in the detector.

We first identify the subspaces ${\cal H}_{\pm}$ that define the transition under consideration. Since the detection proceeds through the
process Eq. (\ref{proc}) and the $A$ particles are not directly
observable, the transition corresponds to subspaces of ${\cal H}_r$. The Hilbert space ${\cal H}_r$
is decomposed as ${\cal H}_0 \oplus {\cal H}_{prod}$, where  ${\cal H}_0$
is the subspace of states prior to the decay $A + B_m \rightarrow
D_n$, and $H_{prod}$ is the subspace corresponding to states of the
decay products. This means that we identify ${\cal H}_-$ with ${\cal H}_{0}$ and ${\cal H}_+$ with ${\cal H}_{prod}$.

We denote as $\hat{P}$ the projector  $\hat{ P}: {\cal H}_r
\rightarrow {\cal H}_{prod}$ and $\hat{Q} = 1 - \hat{P}$ its
complement. The corresponding projectors on ${\cal H}_{tot}$ are $1
\otimes \hat{P}$ and $1 \otimes \hat{Q}$.

 Measurements carried out on the product particles  correspond to a family of positive operators
$1 \otimes \hat{\Pi}_{\lambda}$ onto ${\cal H}_{prod}$, where
$\hat{\Pi}_{\lambda}$ are positive operators on ${\cal H}_r$ corresponding to
different values $\lambda$ of the measured observables. They satisfy the completeness relation $\int d \lambda \hat{\Pi}_{\lambda} = \hat{P}$.

The  Hamiltonian is of the form
\begin{eqnarray}
\hat{H} = \hat{H}_A \otimes \hat{1} + 1\otimes \hat{H}_{r} +
\hat{H}_I,
\end{eqnarray}
where $\hat{H}_A$
 is the Hamiltonian for free  $A$
particles, $\hat{H}_{r}$ is the Hamiltonian for the $B_m$ and $D_n$
particles, and $\hat{H}_I$ is the interaction term.

We assume that any particles $B_m$ that
are present prior to detection are almost stationary. This condition
defines our reference frame, that in most cases coincides with the
laboratory frame. Assuming that the initial state for the $B_m$
particles has support to values of momentum much smaller than their
rest mass, the restriction of the operator $\hat{H}_r$ in the subspace ${\cal
H}_0$ is a constant. We choose this constant so that  the Hamiltonian $\hat{H}_A$ on single-particle states
for the $A$ particles is
\begin{eqnarray}
\hat{H}_A =  \sqrt{m^2 + {\bf \hat{p}}^2} - E_0,
\end{eqnarray}
where ${\bf \hat{p}}$ is the $A$-particle momentum. For simplicity,   we  ignore the A-particles' spin degrees of freedom. The constant $E_0$ is the threshold of the process (\ref{proc})
\begin{eqnarray}
 E_0 = \sum_n M_{D_n} - \sum_m M_{B_m},
 \end{eqnarray}
where $M_{B_m}$ and $M_{D_n}$ are the
masses of the particles $B_m$ and $D_n$ respectively

In the subspace corresponding to the states after the detection events, the Hamiltonian is
\begin{eqnarray}
\hat{H}_r = \sum_n \left(\sqrt{M_{D_n}^2 +
\hat{{\bf p}}_n^2} - M_{D_n} + \hat{V}[\xi(t)] \right), \label{hr}
\end{eqnarray}
 where
  $\hat{{\bf
p}}_n$ are the momentum operators for the $D_n$ particles. As in the model of Sec. 3.1.2 we treat any degrees of freedom of the detector other than the particles involved in the reaction Eq. (\ref{proc}) as an environment, and model their action on the product particles by a stochastic term
  $\hat{V}[\xi(t)]$.

The interaction Hamiltonian is
\begin{eqnarray}
\hat{H}_I = \sum_i \int d^3x \left[\hat{b}({\bf x})
\hat{J}^{+}({\bf x}) + \hat{b}^{\dagger}({\bf x})
 \hat{J}^-({\bf x})\right], \label{hi}
\end{eqnarray}
 where $\hat{b}, \hat{b}^{\dagger}_i$ are annihilation and
creator operators on ${\cal H}_{A}$, and
 $J^{\pm}({\bf x})$ are current operators,    involving products of
annihilation operators for the $B$ particles and creation operators
for the $D$ particles. Since no $A$-particles are created during the
detection process, the initial state $|\phi_0 \rangle$ in ${\cal
H}_r$ must satisfy
\begin{eqnarray}
\hat{J}_{\alpha}^-({\bf x}) |\phi_0 \rangle = 0.
\end{eqnarray}

\medskip

With the definitions above, the time-of-arrival probability, restricted to one dimension, is given by Eqs. (\ref{plt}) and (\ref{fpp}).
 We consider a pointer variable corresponds to the position $X_1$ of  the $D_1$ particle. Thus, the positive operators $\hat{\Pi}_L$ of Eq. (\ref{po}) become

\begin{eqnarray}
\sqrt{\hat{\Pi}_L} = \frac{1}{\sqrt{\pi \delta^2}}\sum_a \int dX_1 \ldots dX_n e^{-\frac{(L-X_1)^2}{2 \delta^2}} |X_1, \ldots, X_n\rangle \langle X_1, \ldots, X_n |, \label{po3}
\end{eqnarray}

Substituting Eq. (\ref{po3}) into Eq. (\ref{fpp}), we obtain terms of the form $\langle X_1, \ldots X_n|\hat{J}^+(x) |\Psi_0\rangle$. The positions of all particles must be close to the value $x$  of the interaction locus, within the accuracy $\delta$ of Eq. (\ref{po2}) corresponding to the localization of the $D_1$ particle. Hence, we  set $X_1 = X_2 = \ldots X_n$ and we express
\begin{eqnarray}
\langle X_1, \ldots X_n|\hat{J}^+(x) |\Psi_0\rangle = \delta(X_1-X_2) \ldots \delta(X_1-X_n) u(X_1 - x),
\end{eqnarray}
 in terms of a function $u(x)$. Then, Eq. (\ref{fpp}) becomes
\begin{eqnarray}
 \langle p'|\hat{S}(L)|p\rangle =  e^{-\delta^2(p-p')^2/4}  \tilde{u}(p) \tilde{u}^*(p') e^{i(pL -ip'L}
 \nonumber \\
\times \int d\tau  e^{-i(\epsilon_p + \epsilon_{p'})\tau/2} \int dZ e^{-\frac{Z^2}{4\delta^2}} e^{i(p+p')Z/2} G(Z, \tau), \label{fpp8}
\end{eqnarray}
 where
 \begin{eqnarray}
G(X-X',s'-s) = {\cal M} \left[ \prod_{n=1}^N  \langle X'|\hat{U}_n(s') \hat{U}^{\dagger}_n(s)| X \rangle \right], \label{ff}
\end{eqnarray}
 In Eq. (\ref{ff}),
$ \hat{U}_n(s) = {\cal T}e^{-i \int_0^s \left(\sqrt{\hat{p}_n^2 +M_n^2} +\hat{V}_n[\xi(s)]\right)}$ stands for the evolution operator for the particle $D_n$, and ${\cal M}$ denotes average over the stochastic process $\xi(\cdot)$ corresponding to the influence of the environment.

Eq. (\ref{fpp8}) gives the time-of-arrival probabilities associated to a general particle reaction (\ref{proc}). The derivation of an explicit formula for a particular detection requires the specification of the field-theoretic interaction (as encoded in the function $u(p)$ and a modeling of the  term
$\hat{V}[\xi(t)]$ that corresponds to the stochastic action of the detector's degrees of freedom on the product particles.

\medskip

Since the diagonal elements of $\hat{S}(L)$ in Eq. (\ref{fpp8}) are independent of $L$, we define an absorption coefficient of the apparatus as
\begin{eqnarray}
\alpha(p) = |\tilde{u}(p)|^2 \int d\tau e^{-i\epsilon_p\tau} \int dZ e^{-\frac{Z^2}{4\delta^2}} e^{ipZ} G(Z, \tau)
\end{eqnarray}

Following the same arguments and approximations as in Sec. 3.1, we show that for initial states narrowly concentrated in momentum,
\begin{eqnarray}
\tilde{P}(L,t) = \left| \int \frac{dp}{2\pi} \sqrt{\alpha(p) |v_p|} \tilde{\psi}_0(p) e^{ipL-i\epsilon_pt}\right|^2,
\end{eqnarray}
that is, we show that Eq. (\ref{plt2b}) for the time-of-arrival probabilities has a broad degree of applicability and its derivation requires no special modeling assumptions.


\section{An application: particle oscillations}

We conclude this paper with a non-trivial application of our formalism in the study of particle oscillations. We give a rigorous measurement-theoretic derivation of the time-integrated probability associated to particle-oscillation experiments. We derive the standard oscillation formula in a regime corresponding to very short decoherence time in the detector, and, intriguingly, we identify a different regime (corresponding to larger decoherence times) that leads to a novel non-standard oscillation formula.

Particle oscillations characterize systems in which the single-particle Hilbert space ${\cal H}_1$ is split into subspaces ${\cal H}_i$ as ${\cal H}_1 = \oplus {\cal H}_i$, such that the dispersion relation $\epsilon_p^i$ is different in each subspace ${\cal H}_i$. In high-energy physics, particle oscillations appear in neutral bosons and in (massive) neutrinos; the different dispersion relations are due to the different values of the mass $m_i$ in the eigenspaces ${\cal H}_i$. Particle oscillations arise because the creation  processes for neutral bosons and neutrinos couple to the particle's flavor, and thus produce superpositions of states with different mass.

To describe particle oscillations, we adapt the formalism of Sec. 3.3.
  We denote the oscillating particles by $A$, and we consider their detection through the channel (\ref{proc}) that corresponds to the flavor $\alpha$. The Hamiltonian on the single-particle Hilbert space ${\cal H}_1$ is of the form
\begin{eqnarray}
\hat{H}_m = \oplus_i \left(\sqrt{\hat{{\bf p}}^2 + m_i^2} - E_0\right),
\end{eqnarray}
where $E_0$ is the threshold of the detection process.

The interaction Hamiltonian is
\begin{eqnarray}
\hat{H}_I = \sum_i \int d^3x \left[\hat{b}_i({\bf x}) U_{ \alpha i}
\hat{J}_{\alpha}^{+}({\bf x}) + \hat{b}^{\dagger}_{i}({\bf x})
U^*_{ \alpha i}\hat{J}^-_{\alpha}({\bf x})\right], \label{hi2}
\end{eqnarray}
 where $\hat{b}_i, \hat{b}^{\dagger}_i$ are annihilation and
creator operators on the Fock space ${\cal H}_{A}$, $i$ labels the mass eigenspaces,
 $J^{\pm}_{\alpha}({\bf x})$ are current operators of flavor
$\alpha$ defined on ${\cal H}_{r}$, and $U_{\alpha i}$ is the mixing
matrix\footnote{A more general treatment (relevant to bosons) would involve a
kernel instead of a constant for the mixing matrix $U_{\alpha i }$,
reflecting the fact that the mixing coefficients may depend on
momentum. Here, however, we shall consider initial states sharply
concentrated in momentum, for which a constant value of $U_{
\alpha i}$ provides a good leading-order approximation.}.

The current operator $\hat{J}_{\alpha}^{\pm}$ in Eq. (\ref{hi2}) involves products of
annihilation operators for the $B$ particles and creation operators
for the $D$ particles. Since no $A$-particles are created during the
detection process, the initial state $|\phi_0 \rangle$ in ${\cal
H}_r$ must satisfy
\begin{eqnarray}
\hat{J}_{\alpha}^-({\bf x}) |\phi_0 \rangle = 0.
\end{eqnarray}

We consider a general single-particle state for the A particles
\begin{eqnarray}
|\psi_0 \rangle =  \sum_i U_{\beta i}^* \int d^3 x \, \hat{b}_{i}^{\dagger}({\bf
x}) \psi_{i0}({\bf x})|0 \rangle \label{psi0}
\end{eqnarray}
where $| 0 \rangle_A$ is the
vacuum of the Fock space ${\cal H}_A$. The dependence the state (\ref{psi0} on the mixing matrix $U^*_{\beta i}$ is a consequence of the  creation of the $A$-particles through an $\beta$-flavor current.

In particle oscillation experiments, the time of arrival of individual neutrinos is not determined. Thus, the relevant quantity is the time-integrated probability density $P_{\beta \alpha}(L)$ for the detection along the $\alpha$-flavor channel of $A$-particles created through the $\beta$-flavor channel. Integrating Eq. (\ref{pp2}) over time $t \in [0, \infty)$, we obtain
\begin{eqnarray}
P_{\beta \alpha}(L) = \sum_{ij}\int_0^{\infty} ds \int_0^{\infty} ds' \int d^3x
d^3x' \psi^*_{i}({\bf x'},s') \psi_{j}({\bf x},s)
 U^*_{\alpha i} U_{ \alpha j} U_{\beta i}^*U_{\beta i} \nonumber \\
 \langle \Psi_0|\hat{J}^-_{\alpha}({\bf x'}) \sqrt{\hat{\Pi}_{L}} {\cal M} \left[ \hat{U}(s')\hat{U}^{\dagger}(s)\right]
  \sqrt{\hat{\Pi}_{L}} \hat{J}^+_{\alpha}({\bf x})|\Psi_0\rangle.
  \label{R1}
\end{eqnarray}
where $\psi_i({\bf x}, t)$ is the Schr\"odinger time evolution of the wave functions $\psi_{i0}({\bf x})$, $\hat{\Pi}_L$ is given by Eq. (\ref{po2}), $\hat{U}(s)$ is the evolution operator for the $D$-particles of Eq. (\ref{proc}) including stochastic terms from interaction with the environment and ${\cal M}$ denotes stochastic averaging. Reducing the system to one spatial dimension and following the procedure that led to the derivation of Eq. (\ref{fpp8}) in Sec. 3.1, we find
\begin{eqnarray}
P_{\beta \alpha}(L) = \sum_{ij}U^*_{\alpha i} U_{ \alpha j} U_{\beta i}^*U_{\beta i}  \int_0^{\infty} ds \int_0^{\infty} ds' \int dp dp' \tilde{\psi}_{j0}(p) \tilde{\psi}_{i0}^*(p') e^{i (p-p')L - i(\epsilon_p^j s - \epsilon^i_{p'}s')} F(p,p',s'-s) \label{probmain}
\end{eqnarray}
in terms of the kernel
\begin{eqnarray}
F(p, p', \tau) = e^{-\delta^2(p-p')^2/4}  \tilde{u}(p) \tilde{u}^*(p') \int dZ e^{-\frac{Z^2}{4\delta^2}} e^{i(p+p')Z/2} G(Z, \tau),
\end{eqnarray}
where $G(Z, \tau)$ is given by Eq. (\ref{ff}). In Eq. (\ref{probmain}), $\epsilon_p^i = \sqrt{p^2+m_i^2} - E_0$.

Next, we  consider a broad class of initial states
\begin{eqnarray}
\psi_{i0}(x) = \phi_0(x)e^{i \bar{p}_i x},
\end{eqnarray}
where $\phi_0$ is some real-valued wave function centered around $x = 0$  with position spread $\sigma_x$ and $\bar{p}_i$ is the mean momentum of the state in the $i$-th mass eigenspace. We assume that $|\bar{p}_i - \bar{p}_j|<< |\bar{p}_i|$ for all $i$ and $j$.

In the momentum representation,
\begin{eqnarray}
\tilde{\psi}_{i0}(p) = \tilde{\phi}_0(p - \bar{p}_i),
\end{eqnarray}
where $\tilde{\phi_0}(p)$ is the Fourier transform of $\phi_0(x)$.

 In order to perform the integration over $p$ and $p'$ in Eq. (\ref{probmain}), we expand
\begin{eqnarray}
\epsilon_p = \bar{\epsilon}_i + \bar{v}_i (p - \bar{p}_i),
\end{eqnarray}
where $\bar{\epsilon}_i = \epsilon_{\bar{p}_i}$, and $\bar{v}_i = (\partial \epsilon^i_p/\partial p)_{p = \bar{p}_i}$. We further assume that the variation of the kernel $F(p,p', \tau)$ in the range of values of momenta where $\tilde{\psi}_{i0}$ is supported is negligible; hence,  $F(p,p',\tau)$ can be treated as a constant in the integration over $p$ and $p'$ in Eq. (\ref{probmain}).
Then, we obtain
\begin{eqnarray}
P_{\beta \alpha}(L) = \sum_{ij}U^*_{\alpha i} U_{ \alpha j} U_{\beta i}^*U_{\beta i}  e^{i(\bar{p}_j - \bar{p}_i)L}
\int_0^{\infty} ds \int_0^{\infty} ds' \phi_0(L-v_js) \phi_0(L-v_is') e^{ - i (\bar{\epsilon}_js - \bar{\epsilon}_is')} f(s'-s) \label{pab2}
\end{eqnarray}
In Eq. (\ref{pab2}), we denoted $f(\tau) = F(\bar{p},\bar{p}, \tau)$, where $\bar{p}$ is the mean momentum of the state $|\psi_0 \rangle$.

\paragraph{The standard oscillation formula.}
The integrand in Eq. (\ref{pab2}) is peaked around $s = L/\bar{v}$ and $s' = L/\bar{v}_i$, namely, around the classical values of the time of arrival corresponding to the dispersion relation in the subspace ${\cal H}_i$. The probability density Eq. (\ref{pab2}) is strongly sensitive on the form of the function $f(\tau)$, which determines whether the amplitudes associated to different times of detection contribute coherently in the total probability. In general, the function $f(\tau)$ depends on the internal dynamics of the detector and on the particles' energy scale. It is expected to vanish for sufficiently large values of $\tau$. For example, if one includes the coarse-graining time-scale $\sigma$ in the derivation, according to Eq (\ref{pp}), $f(\tau)$
would include a multiplicative Gaussian term $\exp(-\tau^2/\sigma^2)$ and it would tend to zero for $\tau >> \sigma$. In general, the presence of incoherent interactions (as in the model of Sec. 3.1.2) implies that $f_{\tau}$ is characterized by some time-scale $\tau_{dec} < \sigma$, such that $f(\tau) \simeq 0$ for
 $\tau >> \tau_{dec}$.

The decoherence time scale $\tau_{dec}$ depends on the physics of the detector and it cannot be specified without a precise modeling of the associated interactions, including the effects of the environment.It plays a crucial role to the form of the probability density (\ref{pab2}). If $\tau_{dec} << |L/\bar{v}_i - L/\bar{v_j}|$, amplitudes peaked at different values of the time of arrival do not contribute coherently to the total probability. In that case, the function $f(s'-s)$ in Eq. (\ref{pab2}) is effectively proportional to a delta function. It follows that the probability density $P_{\beta \alpha}$ of Eq. (\ref{pab2}), is proportional to $\int_0^{\infty} ds |\psi(L,s)|^2$, as it is assumed in the so-called wave-packet description of particle oscillations \cite{wavepacket, equaltime}.

In this regime, the evaluation of the probability Eq. (\ref{pab2}) involves the integral
\begin{eqnarray}
\int_{-\infty}^{\infty} ds \phi_0(L-\bar{v}_is)  \phi_0(L-\bar{v}_js) e^{ - i (\bar{\epsilon}_j - \bar{\epsilon}_i)s}, \label{int1}
\end{eqnarray}
where we extended the limits of integration to $(-\infty, \infty)$, since for $L >> \sigma_x$ the integrand is strongly suppressed for values $s < 0$.

The integral Eq. (\ref{int1}) is to be estimated subject to the condition $|\bar{v}_i - \bar{v}|<<\bar{v}_i$, for all $i$ and $j$. To this end, we change the integration variable to $r = - s + \frac{L}{\bar{v}}$, where $\bar{v}$ is a mean velocity in the initial state $|\psi_0 \rangle$ \footnote{For example, $\bar{v}$ may be defined as the arithmetic mean or as the geometric mean of $\bar{v}_i$ and $\bar{v}_j$. As long as $|\bar{v}_i - \bar{v}_j| << \bar{v}$, the results are not affected by the way we choose to define the mean velocity.}. The integral Eq. (\ref{int1}), then becomes
\begin{eqnarray}
e^{ - i (\bar{\epsilon}_j - \bar{\epsilon}_i)L/\bar{v}} \int_{-\infty}^{\infty} dr \phi_0(-\frac{\delta v_i}{\bar{v}}L +\bar{v}_ir)  \phi_0(-\frac{\delta v_i}{\bar{v}}L + \bar{v}_jr) e^{  i (\bar{\epsilon}_j - \bar{\epsilon}_i)r}, \label{int2}
\end{eqnarray}
where $\delta v_i = \bar{v}_i - \bar{v}$. By Fourier-transforming the functions $\phi_0$ in Eq. (\ref{int2}), we can estimate the leading-order contribution to the integral as $\phi_1 \left(\frac{\bar{v}_i -\bar{v}_j}{\bar{v}} L\right)$, where $\phi_1(x)$ is the inverse Fourier transform of $|\tilde{\phi}(p)|^2$. The spread of $\phi_1(x)$ is of the same order as $\sigma_x$. Therefore, if
\begin{eqnarray}
L << L_{loc}:= \sigma_x \bar{v}/(\bar{v}_i - \bar{v}_j),
 \end{eqnarray}
 the term $\phi_1 \left(\frac{\bar{v}_i -\bar{v}_j}{\bar{v}} L\right)$ is approximately a constant. The parameter $L_{loc}$ is known as the {\em localization length}; if $L>> L_{loc}$ the detection probability is strongly suppressed \cite{wavepacket, Beuth}.

In the regime where $L << L_{loc}$, Eq. (\ref{pab2}) becomes
\begin{eqnarray}
P_{\beta \alpha}(L) =   \sum_{ij}C^1_{ij} U^*_{\alpha i} U_{ \alpha j} U_{\beta i}^*U_{\beta i} e^{i(\bar{p}_j - \bar{p}_i)L - i (\bar{\epsilon}_j - \bar{\epsilon}_i)L/\bar{v}}, \label{standard}
\end{eqnarray}
where  $C^1_{ij}$ are positive constants.

The probability density Eq. (\ref{standard}) is a periodic function of the distance $L$ of the detector from the source with oscillation wave-numbers
\begin{eqnarray}
k_{ji} = (\bar{p}_j - \bar{p}_i) - \frac{1}{\bar{v}}(\bar{\epsilon}_j - \bar{\epsilon}_i) \label{kijst}
\end{eqnarray}
In general, the value of $k_{ij}$ in Eq. (\ref{kijst}) depends on the mean values of the momenta  $\bar{p}_i$ on the subspaces ${\cal H}_i$. Since the interactions that produce the oscillations couple to the flavor basis, there is no reason for the value of momentum in a given subspace to be consistently larger than the momentum in another subspace; hence, we expect that when averaging over the ensemble $\bar{p}_i = \bar{p}_j = \bar{p}$. Different assumptions about the initial state have been employed in the literature \cite{equaltime, Beuth}, for example, that the energies $\bar{\epsilon}_i$ are equal.  In the context of the present formalism,  such assumptions do not affect the resulting probability distributions.

Substituting for all $\bar{p}_i$ the mean momentum $\bar{p}$ of the initial state, we obtain
\begin{eqnarray}
k_{ji} = \frac{m_i^2 - m_j^2}{2 \bar{p}}, \label{kijst3}
\end{eqnarray}
and we recover the standard expression for the oscillation wavenumber that applies both to neutral-boson and neutrino oscillations.

\paragraph{A non-standard oscillation formula.}
The standard oscillation formula, Eq. (\ref{kijst3}),  was obtained from the assumption that $\tau_{dec} << |L/\bar{v}_i - L/\bar{v}_j|$. In the opposite regime, where the decoherence time is sufficiently large so that $\tau_{dec} >> |L/\bar{v}_i - L/\bar{v}_j|$, amplitudes peaked at different values of  the time of arrival contribute coherently in Eq. (\ref{pab2}). In this regime, the function $f(s'-s)$ is essentially constant and equal to $f(0)$. Hence,
\begin{eqnarray}
P_{\beta \alpha}(L) =  \sum_{ij}C^2_{ij} U^*_{\alpha i} U_{ \alpha j} U_{\beta i}^*U_{\beta i} e^{i(\bar{p}_j - \bar{p}_i)L - i (\frac{\bar{\epsilon}_j}{\bar{v}_j} - \frac{\bar{\epsilon}_i}{\bar{v}_i})L}, \label{nonstandard}
\end{eqnarray}
where $C^2_{ij}$  are positive constants. In this regime, there is no coherence length $L_{coh}$.

 The probability density Eq. (\ref{nonstandard}) is a periodic function of the distance $L$ of the detector from the source with oscillation wave-numbers
\begin{eqnarray}
k_{ji} = (\bar{p}_j - \bar{p}_i) -  (\frac{\bar{\epsilon}_j}{\bar{v}_j} - \frac{\bar{\epsilon}_i}{\bar{v}_i}) = \frac{m_i^2}{\bar{p}_i} - \frac{m_j^2}{\bar{p}_j} +E_0\left(\frac{1}{\bar{v}_i}-\frac{1}{\bar{v}_j}\right). \label{kijnst}
\end{eqnarray}

Eq. (\ref{kijnst}) does not depend on the precise choice of the mean momenta $\bar{p}_i$. Again, setting $\bar{p}_i = \bar{p}_j = \bar{p}$, Eq. (\ref{kijnst}) becomes
\begin{eqnarray}
k_{ji} = \frac{m_i^2 - m_j^2}{\bar{p}} - \frac{E_0}{ \bar{p}} \left[\sqrt{m_i^2+\bar{p}^2} - \sqrt{m_j^2+\bar{p}^2}\right].
\label{kijnst2}
\end{eqnarray}
 Eq. (\ref{kijnst2}) is a non-standard oscillation formula. Its dependence on the threshold energy $E_0$ is particularly notable, because it implies a different oscillation wavelengths for different channels of detection.

For $E_0 = 0$, Eq. (\ref{kijnst2}) becomes
\begin{eqnarray}
k_{ij} = (m_i^2 - m_j^2)/\bar{p},
 \end{eqnarray}
 i.e., it predicts an oscillation wavenumber twice as large  as the standard expression. For neutrinos, this non-standard oscillation formula has been derived through other methods---see Ref. \cite{nst} and Refs. \cite{equaltime, cri} for critique. In absence of independent measurements of the mass differences $m_i - m_j$, the nonstandard oscillation formula is indistinguishable experimentally from the standard one.

For $E_0 >0$, we consider separately the non-relativistic regime (relevant to neutral bosons) and the ultra-relativistic regime relevant to neutrinos.

 In the non-relativistic regime, where $|m_i - m_j| << m_i$, we define a "mean mass" $m:= \frac{m_i+m_j}{2}$, to obtain
\begin{eqnarray}
k_{ji} = \frac{m_i - m_j}{\bar{p}} (2 m -E_0)  \label{kijnstnr}
\end{eqnarray}
Eq. (\ref{kijnstnr}) has the same dependence on momentum as the standard expression for the oscillation wavelength, and thus, it is indistinguishable in absence of an independent measurement of the mass differences $m_i - m_j$.

In the ultra- relativistic regime,
\begin{eqnarray}
k_{ji} = \frac{m_i^2 - m_j^2}{2\bar{p}} \left( 1 - \frac{E_0}{2\bar{p}}\right). \label{kijnstur}
\end{eqnarray}
The momentum dependence of $k_{ji}$ in Eq. (\ref{kijnstur}) differs from that in the standard oscillation formula. The difference is more pronounced as $\bar{p}$ approaches the threshold energy $E_0$.

To summarize, the derivation of the standard expression for particle oscillations requires the assumption that `interferences' in the time of arrival are decohered at the detector. In absence of a strong decoherence effect, the virtual processes, peaked around different values for the time of arrival, contribute coherently to the total probability. They result to a different expression of the oscillation wavelength. For a given detection process, the standard oscillation formula applies for sufficiently large values of $L$, and the non-standard oscillation formula applies for sufficiently small values of $L$. In  typical neutrino oscillations the baseline $L$ is of the order of $10^2 m$ and the neutrino energies of the order of hundreds of $MeVs$; then
 $|L/\bar{v}_i - L/\bar{v_j}|$ is of the order of $10^{-22}s$. The validity of the standard oscillation formula requires that the decoherence time-scale be much smaller than $10^{-22}s$. This time-scale is very small, and at the moment, there exists no first-principles modeling of actual detectors to establish whether this value is physically realistic or not. For this reason, we believe that there is a good {\em prima facie} possibility that  the non-standard oscillation formulas could be physically relevant \cite{AnSav10}.

\section{Conclusions}

The main contribution of this paper is the development of an   method for determining the time-of-arrival probabilities, valid for to any experimental set-up. Our method is algorithmic. in the sense that for any modeling of the detector that determines the particle, a unique expression for the time-of-arrival probability follows. The method is also general, because it can incorporate any interaction between microscopic system and detector, including ones described in terms of relativistic quantum field theory. This achieved this result by: (i)    reducing the problem of defining quantum temporal observables to a mathematical model where time is associated to a transition from a subspace of the Hilbert space of the  system to its complementary subspace, and (ii) combining a quasi- classical description of the measurement records with a fully quantum modeling of the detector's interaction with the microscopic system.

We constructed time-of-arrival probabilities for three different detector models. We showed that there exists a special regime in which all information about the detector is encoded in a single function of momentum, the absorption coefficient $\alpha(p)$. From these models, we obtain a generalization of Kijowski's probability distribution, valid for any dispersion relation for the microscopic particle. We also adapted our formalism for the description of the time-of-arrival in high-energy physics, where the detection process involves the creation of several product particles on the detector. As a non-trivial application of the method, we constructed rigorously the time-integrated probability associated to particle-oscillation experiments.

The applicability of the method is not restricted to the time-of-arrival problem. It can be employed in order to define probabilities associated to any physical transition (for example, decays of unstable systems), provided that such a transition is accompanied by a macroscopic record of observation. Furthermore, it applies to set-ups involving more than one detectors \cite{AnSav11}. Thus, it can be employed for the construction of temporal correlation functions between different detectors, or in order to construct temporal entanglement witnesses associated to measurements in multi-partite systems. Furthermore, since the method is compatible with relativistic quantum field theory, it can also be employed towards the construction of quantitatively precise models of relativistic quantum measurements \cite{PT}, in which the spacetime coordinates of an event are random variables rather than externally predetermined parameters.

\end{document}